\documentclass{article}

\usepackage{arxiv}

\usepackage[utf8]{inputenc} 
\usepackage[T1]{fontenc}    
\usepackage{fancyhdr}
\usepackage{lmodern}
\usepackage[colorlinks]{hyperref}  
\usepackage{url}            
\usepackage{booktabs}       
\usepackage{amsfonts}       
\usepackage{nicefrac}       
\usepackage{microtype}      
\usepackage{lipsum}

\usepackage{graphicx}
\usepackage{bm}
\usepackage{epsfig}
\usepackage{epstopdf}
\usepackage{subfigure}

\usepackage{enumerate}
\usepackage{amssymb}
\usepackage{amsmath}
\usepackage{amsthm}
\usepackage{enumitem}
\usepackage{mathrsfs}

\usepackage{multirow}%
\usepackage[title]{appendix}%
\usepackage{xcolor}%
\usepackage{textcomp}%
\usepackage{manyfoot}%
\usepackage{algorithm}%
\usepackage{algorithmicx}%
\usepackage{algpseudocode}%

\renewcommand{\bar}{\overline}
\renewcommand{\rho}{\varrho}
\renewcommand{\phi}{\varphi}

\theoremstyle{plain}
\theoremstyle{definition}

\title{Turing chemotactic instability in an HIV model}

\author{  
Florinda Capone\href{https://orcid.org/0000-0002-0672-999X}{\includegraphics[scale=0.1]{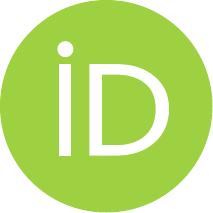}} \\ Dipartimento di Matematica e Applicazioni ``R.Caccioppoli'' \\ Universit\'a degli Studi di Napoli Federico II \\ Via Cintia, Monte S.Angelo, 80126 Napoli \\ Italy \\ 
\texttt{fcapone@unina.it} \\
\And
Roberta De Luca\thanks{Corresponding author.} \href{https://orcid.org/0000-0002-2109-7564}{\includegraphics[scale=0.1]{orcid}} \\ Dipartimento di Matematica e Applicazioni ``R.Caccioppoli'' \\ Universit\'a degli Studi di Napoli Federico II \\ Via Cintia, Monte S.Angelo, 80126 Napoli \\ Italy \\   
\texttt{roberta.deluca@unina.it} \\ 
\And
Vincenzo Luongo\href{https://orcid.org/0000-0001-8875-4963}{\includegraphics[scale=0.1]{orcid}} \\ Dipartimento di Matematica e Applicazioni ``R.Caccioppoli'' \\ Universit\'a degli Studi di Napoli Federico II \\ Via Cintia, Monte S.Angelo, 80126 Napoli \\ Italy \\   
\texttt{vincenzo.luongo@unina.it}
} 

\renewcommand{\shorttitle}{Turing chemotactic instability in an HIV model}

\begin{document}
\maketitle

\begin{abstract}

A ternary reaction-diffusion model for early HIV infection dynamics, incorporating logistic growth of target cells, is introduced. According to \emph{in vitro} and \emph{in vivo} studies, random movement of target cells, infected cells, and virions and a chemotactic attraction of target cells by cytokines, are included. The research explores the existence of disease-free and coexistence equilibria, conducting linear stability analyses for homogeneous and heterogeneous scenarios.
Specifically, conditions for chemotactic-self diffusion instability of the endemic equilibrium are found, indicating that Turing patterns may emerge when the chemotactic effect surpasses a critical threshold. This threshold is lower than in models without logistic growth of target cells, suggesting that the logistic model provides better insights into infection hot spots in the early stages. The location and shape of these patterns, crucial for developing infection control strategies, are investigated using weakly nonlinear analysis and demonstrated through numerical simulations.
\end{abstract}

\keywords{HIV model \and Turing instability \and Chemotaxis \and Reaction-Diffusion}

\section{Introduction}

The Human Immunodeficiency Virus (HIV) is a retrovirus that, untreated, can lead to AIDS, characterized by a severe weakening of the immune system.  Despite advancements in treatment, HIV remains a significant global health concern with millions affected and hundreds of thousands of deaths annually (see https://www.who.int/news-room/fact-sheets/detail/hiv-aids).
The mechanism of infection is as follows: the HIV virus primarily targets a specific type of white blood cell known as T-helper cells, also referred to as CD4$^+$ T-lymphocytes, as well as monocytes/macrophages, key components of the immune system. Upon infecting these cells, the virus converts its RNA into DNA, which integrates into the host cell's DNA. This integration allows the virus to enter a latent state, evading immune detection. Alternatively, the virus can undergo transcription to produce new viral RNA and proteins, which assemble into new virus particles released from the cell, restarting the infection cycle \cite{Belasio}.

Mathematical models, predominantly ODEs, have been crucial in understanding HIV dynamics. These models typically describe viral replication dynamics and the immune response but often overlook spatial variations. In \cite{Nowak}, \cite{Perelson96}, \cite{Wei95}  compartmental models are introduced combining the fundamental SI-model for CD4$^+$ T cells, divided into sound (or target) cells and productively infected cells, with virions dynamic. In \cite{Perelson99}, a model incorporating viral production for HIV-$1$ infection is introduced. In particular, the dynamics through which the infected CD4$^+$ T cells can produce new HIV virus particles is modeled. 

Turing instability, first theorized in \cite{Turing}, explores how diffusion can destabilize a steady state, leading to spatial patterns in populations.
In the context of HIV modeling, Stancevic \cite{Stancevic} introduced a reaction-diffusion model focusing on early infection dynamics, incorporating self-diffusion and chemotaxis. This work identifies conditions for Turing instability, revealing spatial hotspots of infection through numerical simulations.

This paper extends the analysis from Stancevic \cite{Stancevic} by focusing on several objectives: identifying necessary and sufficient conditions for Turing instability onset, examining the shape and stability of emerging Turing patterns more comprehensively, and generalizing findings to include realistic logistic growth of T cells.

Our study reveals that the model dynamics incorporating logistic growth of T cells align more closely with real-world scenarios. Specifically, we find that the threshold for Turing instability is lower in our model compared to that in Stancevic \cite{Stancevic}, suggesting better predictive capability for early infection hot spots. Additionally, we conduct a weakly nonlinear analysis to deepen our understanding of pattern formation, exploring their amplitude and shape.

The paper is structured as follows. Section 2 discusses key biological aspects essential for mathematical modeling of HIV infection, aiming for a biologically accurate model to predict early disease evolution \cite{Haase}. Section 3 presents a mathematical model describing interactions among target (T) and infected (I) CD4+ T cells, and virions (V), extending Stancevic's model \cite{Stancevic} to include logistic growth of T cells. It analyzes the disease-free and endemic equilibria and investigates the impact of new parameters such as intrinsic growth rate and carrying capacity of T cells. Section 4 conducts preliminary numerical analyses to explore solution dynamics concerning the chemotactic coefficient, demonstrating that exceeding a specific threshold leads to spatial pattern formation. Sections 5 and 6 perform linear stability analyses of the equilibria. In particular, Section 5 examines stability without spatial variation, while Section 6 incorporates diffusion and chemotaxis, deriving closed-form necessary and sufficient conditions for Turing instability. Section 7 conducts weakly nonlinear analysis near the bifurcation point to understand pattern formation dynamics, including amplitude determination. Section 8 provides a recap of the obtained results, highlighting the contributions and implications of the study.

Overall, this work advances the understanding of HIV infection dynamics through rigorous mathematical analysis, offering insights into pattern formation crucial for disease management strategies.

\section{Key biological aspects and model parameters}

Key biological aspects essential for formulating the mathematical model are the following.
T-cells, produced by sources such as the thymus, are supplied at a constant rate $s$, which ranges between $0$ and $10$ $cells\, mm^{-3}\, day^{-1}$ \cite{Nelson}. CD4+ T-cell counts are crucial for understanding HIV infection stages. Healthy individuals typically have $500$ to $1500$ $cells\, mm^{-3}$ \cite{Lai2015modeling}, with counts below $200$ $cells\, mm^{-3}$ indicating AIDS \cite{Battistini}. The proliferation of T-cells follows a logistic growth rate, dependent on cell density, with intrinsic growth rate $\alpha$ and carrying capacity $k$ \cite{Ho, Sach}. The infection rate, proportional to the product of target cells and virus, follows a mass-action form, with the contact rate $\beta$ estimated in \cite{Ho}. Mortality rates for T and I cells are $\mu_T$ and $\mu_I$, respectively, and the virion clearance rate is $\mu_V$ \cite{Perelson96}. Infected cells produce virions at a rate $\gamma=N\mu_I$, with $N$ being the number of virions produced per infected cell. Early-stage HIV infection often presents with a rash, and microscopic infection hotspots are observed \cite{Haase}.
To account for spatial phenomena, the model includes random cell movement, enabling the study of Turing instability and spatial hotspots of infection. T-cells move through the bloodstream and lymphatic system, with experimental studies demonstrating self-diffusion \cite{Miller2003}. Diffusion coefficients for T, I cells, and virions $(D_T,D_I,D_V)$ have been estimated \cite{Miller2002, Stancevic}. Table 1 provides values for all biological parameters used in the numerical study (see Section 3).

\begin{table}
\centering
\begin{tabular}{|c|c|}
  \hline
   $\beta= 3.43\cdot 10^{-5}$ ml virions$^{-1}$ day$^{-1}$ & $s=10$ cells mm$^{-3}$ day$^{-1}$\\
  \hline
  $\mu_I= 0.5 $ day$^{-1} $ & $\mu_V=$ 3 day$^{-1}$\\
  \hline
  $\mu_T=$ 0.03 day$^{-1} $ & $D_T$= 0.09504 mm$^2$ day$^{-1}$\\
  \hline
  $D_I=$ 0.085 mm$^2$ day$^{-1}$ & $D_V=$ 0.0076032 mm$^2$ day$^{-1}$\\
  \hline
   \end{tabular}
\caption{Estimated values of the parameters in (\ref{model})}
\end{table}

\section{Mathematical model and equilibria}

The mathematical model governing the evolution of $T,I,V$ in the spatial regular domain $\Omega\subset \mathbb R^3$ is \cite{Stancevic}, \cite{Perelson92}
\begin{equation}\label{model}
\begin{cases}
  T_{,\tau}=s-\mu_TT+\alpha T\left(1-\dfrac{T}{k}\right)-\beta VT+D_T\Delta T-\chi\nabla\cdot(T\nabla I),\\
  I_{,\tau}=\beta VT-\mu_I I+D_I\Delta I,\\
  V_{,\tau}=\gamma I-\mu_VV+D_V\Delta V,
\end{cases}
\end{equation}
being $\varphi_{,\tau}$ the partial derivative of $\varphi(\mathbf x^*,\tau)$ with respect to time $\tau$; $\nabla(\cdot)$, $\Delta(\cdot)$ and $\nabla\cdot (\cdot)$, the nabla, Laplacian and divergence operator with respect to the space variable $\mathbf x^*$. To (\ref{model}) we append smooth positive initial data and no-flux boundary conditions, i.e.,
\begin{equation}\label{inboucond}
\varphi(\mathbf x^*, 0)=\varphi_0(\mathbf x^*)>0,\qquad \nabla\varphi\cdot\mathbf n=0,\,\mbox{ on }\partial \Omega\times \mathbb R^+,\,\,\forall\varphi\in\{T,I,V\}.
\end{equation}
All the parameters in (\ref{model}) are supposed to be positive. The biologically meaningful constant equilibria are
\begin{itemize}
 \item[1)] the \emph{disease-free equilibrium} $E_0=(\bar T, 0,0)$ with
\begin{equation}\label{Tdfe}
 \bar T=\dfrac{k(\alpha-\mu_T)+\sqrt{k^2(\alpha-\mu_T)^2+4ks\alpha}}
 {2\alpha};
 \end{equation}
 \item[2)] the \emph{endemic equilibrium} $E_c=(T_c,I_c, V_c)$
 with 
 \begin{equation}\label{endequi}
\!\!\!\! \begin{array}{l}
 T_c\!=\!\dfrac{\mu_I\mu_V}{\beta \gamma},\, V_c\!=\!\dfrac{\gamma I_c}{\mu_V},\,
 I_c\!=\!\dfrac{sk\beta^2\gamma^2+ k\mu_I\mu_V(\alpha-\mu_T)\beta \gamma-\alpha \mu^2_I\mu^2_V}{\beta^2\gamma^2 k\mu_I}
\end{array}\end{equation}
existing only if 
$
\beta\gamma>\dfrac{\mu_I\mu_V}{2sk}\left[k(\mu_T-\alpha)+\sqrt{k^2(\mu_T-\alpha)^2+4\alpha s k}\right].
$
 \end{itemize}

In order to obtain general results holding independently of the units, let us introduce the no-dimensional transformation
\begin{equation}\label{adim}
u=T/T_c,\quad v=I/I_c; \quad w=V/V_c,\quad \mathbf x=\mathbf x^*/L,\quad t=\mu_T \tau,\end{equation}
being $L$ the $\Omega$ diameter.
Setting
\begin{equation}\label{paradi}
\begin{array}{l}
 \xi=\dfrac{s}{\mu_T T_c},\,\, \delta=\dfrac{\alpha}{\mu_T},\,\, h=\dfrac{k}{T_c},\,\, \eta=\dfrac{\beta\gamma}{\mu_T\mu_V},\,\,\mu_1=\dfrac{\mu_I}{\mu_T},\,\, \mu_2=\dfrac{\mu_V}{\mu_T},\\
 d_1=\dfrac{D_T}{L^2\mu_T},\,\,d_2=\dfrac{D_I}{L^2\mu_T},\,\,
 d_3=\dfrac{D_V}{L^2\mu_T},\,\,d=\dfrac{\chi}{L^2\mu_T},
\end{array}\end{equation}
the no-dimensional system becomes
\begin{equation}\label{adimode}
\begin{cases}
 \dfrac{\partial u}{\partial t}=\xi-u+\delta u\left(1-\dfrac{u}{h}\right)-\eta I_c u w+d_1\Delta u-dI_c\nabla\cdot(u\nabla v),\\
 \dfrac{\partial v}{\partial t}=\mu_1(uw-v)+d_2\Delta v,\\
 \dfrac{\partial w}{\partial t}=\mu_2(v-w)+d_3\Delta w
\end{cases}\end{equation}
under the initial and boundary conditions 
\begin{equation}\label{ibc}
\varphi(\mathbf x,0)=\varphi_0(\mathbf x), \qquad \nabla\varphi\cdot\mathbf n=0, \ \ \mbox{on}\partial\Omega\times\mathbb R^+,\forall\varphi\in\{u,v,w\}.\end{equation}
Then, the equilibria become $E_0=(\bar u,0,0)$ with

$\bar u=\dfrac{h(\delta-1)+\sqrt{h^2(\delta-1)^2+4h\delta\xi}}{2\delta},$
and $E_c=(1,1,1)$ existing if and only if
\begin{equation}\label{exist}
\xi-1+\delta\dfrac{(h-1)}{h}>0.\end{equation}
Denoting by $R_0=\xi+\delta\dfrac{(h-1)}{h}$ the basic reproduction number, condition (\ref{exist}) is equivalent to $R_0>1$. Let us remark that, in view of (\ref{paradi})$_1$, (\ref{paradi})$_3$ and the biological restriction $s<k\mu_T$, it is not possible that $\xi>1$ and $h<1$. In fact, $\xi>1$ is equivalent to require that $s>T_c\mu_T$. Then one has $T_c\mu_T<s<k\mu_T$ that is impossible  when $h<1$ (i.e. when $k<T_C$). The condition (\ref{exist}) implies a restriction on the parameters $\delta$ and $h$. Precisely:
\begin{itemize}
\item when $\xi<1$ (this happens, for example, when the production rate of $T$ cells is low), (\ref{exist}) is always verified only when the no-dimensional carrying capacity and intrinsic growth rate, are sufficiently high. In particular, in the case $\xi<1$, (\ref{exist}) is verified if and only if
\begin{equation} 
h>1,\qquad \delta>\frac{h(1-\xi)}{h-1};\end{equation}
\item when $\xi>1$ then (\ref{exist}) is verified if and only if $h>1$.
\end{itemize}
Then, a necessary condition for the existence of $E_c$ is that the no-dimensional carrying capacity is $h>1$. We remark that, from (\ref{endequi})$_3$ and (\ref{paradi}), it easily follows that
\begin{equation}\label{12*}
\eta I_c= \xi-1+\delta-\dfrac{\delta}{h}.\end{equation}

\section{Preliminary numerical analysis}

Before starting with the qualitative analysis of the model, we perform numerical simulations to deeply explore the evolution of the infection. In particular, we analyze simulations in the bi-dimensional spatial case.

First of all we recall that, in two spatial dimensions, we need a regularization term to overcome the unrealistic phenomenon of the blow up of solutions in a finite time if the chemotaxis exceeds a certain threshold (see \cite{Hillen} and the references therein for more details). According to \cite{Stancevic} and \cite{Velasquez}, we introduce a density-dependent sensitivity
regularization by substituting the chemotactic term in (\ref{adimode})$_1$ with
\begin{equation}
\label{regchem}
-(1+\varepsilon) d I_c \nabla\cdot\left(\dfrac{u}{1+\varepsilon u}\nabla v\right), \qquad \varepsilon\geq0.
\end{equation}
\noindent
Introducing the effective chemotaxis $\tilde d(u,\varepsilon)=\dfrac{1+\varepsilon}{1+\varepsilon u}d,$ one recovers that, at the endemic equilibrium, $\tilde d=d$ and hence all the results obtained for the linear stability of regularized model continue to hold for the linear stability analysis of the endemic state of the original model.

Numerical simulations have been performed using \emph{Comsol Multiphysics} platform. The non-dimensional version of the mathematical model was integrated by using a built-in highly nonlinear solver and using an extremely fine spatial mesh grid. According to \cite{Stancevic}, the special domain was set to reproduce a tissue thickness of $0.10 \,mm$ defined in a squared domain of $L=1.78\, mm$.

Preliminary numerical simulations were carried out to study the effect of the newly introduced biological parameters $\alpha$ and $k$. In the case of the logistic growth rate, different values were tested from $0.005$ to $3\, day^{-1}$. Moreover, the carrying capacity constant $k$ was varied within the biologically meaningful range of $500$ and $1500$ $cell\, mm^{-3}$ \cite{Lai2015modeling}. The pattern was effectively influenced by the logistic rate when $\alpha$ was set at the same magnitude of the target cells suppling term $s$. Turing patterns were observed in all numerical simulations showing a faster definition by increasing the carrying capacity $k$ and the logistic growth rate $\alpha$. Values of 240, 480, and 960 virions per cell were tested  for $N$ using the same effective chemotaxis constant. The simulations showed that a small $N$ led to lower concentrations of the $w$ population, resulting in a slower infection rate. Since virion concentration $w$ has a relatively slow infection effect on cells, the value of 960 was chosen in line with previous work \cite{Stancevic}. The chemotaxis regularization term $\varepsilon$, was varied from 0 to 50. As expected, a solution blow-up occurred for very short simulation times with $\varepsilon=0$. No significant variations were observed for $\varepsilon\geq 15$, so the regularization term was set to 15 or 25 in subsequent simulations.

The first numerical application demonstrated the stability properties of the model solution under limited chemotactic effect. The non-dimensional chemotaxis coefficient was set around $9.5\, mm^4\, cell^{-1}\, day^{-1}$. Using the endemic equilibrium as the initial condition and applying random perturbations to the infected cells $v$ in the spatial domain, all species exhibited common oscillating behavior before returning to the unperturbed endemic equilibrium.

\begin{figure}[!htbp]
    \centering
    \begin{tabular}{ccc}
        \includegraphics[width=0.2\linewidth]{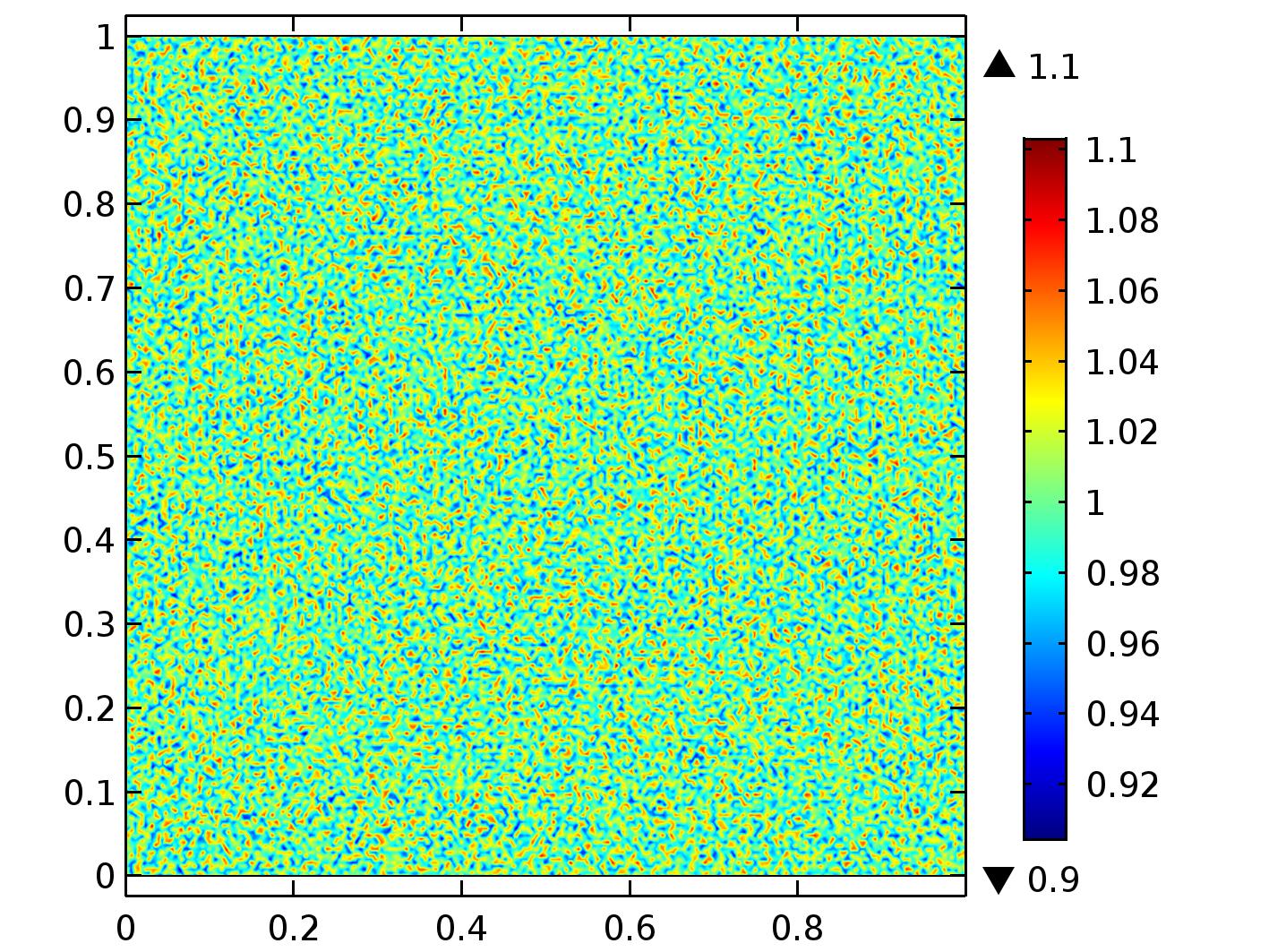} &
        \includegraphics[width=0.2\linewidth]{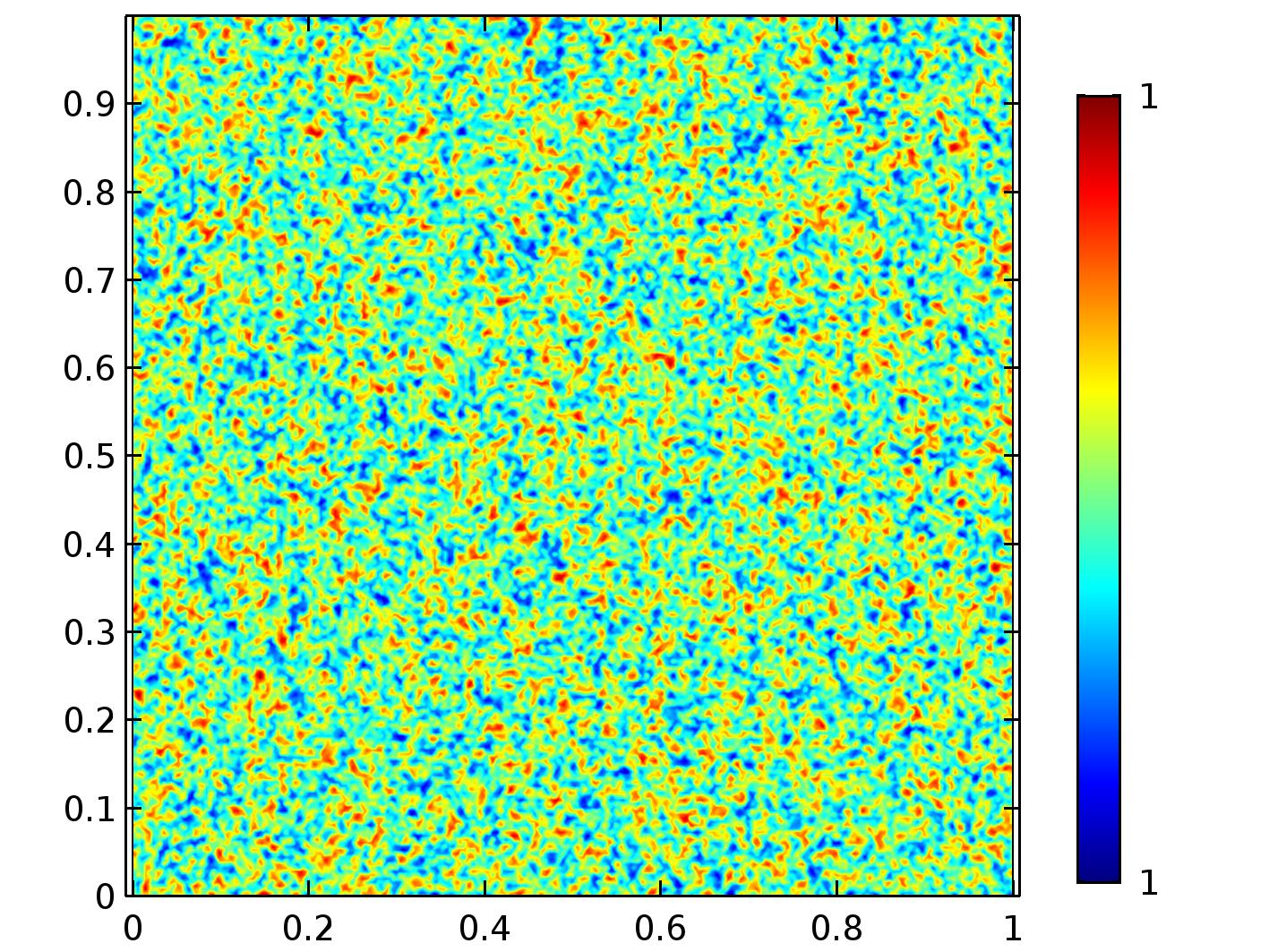} &
        \includegraphics[width=0.2\linewidth]{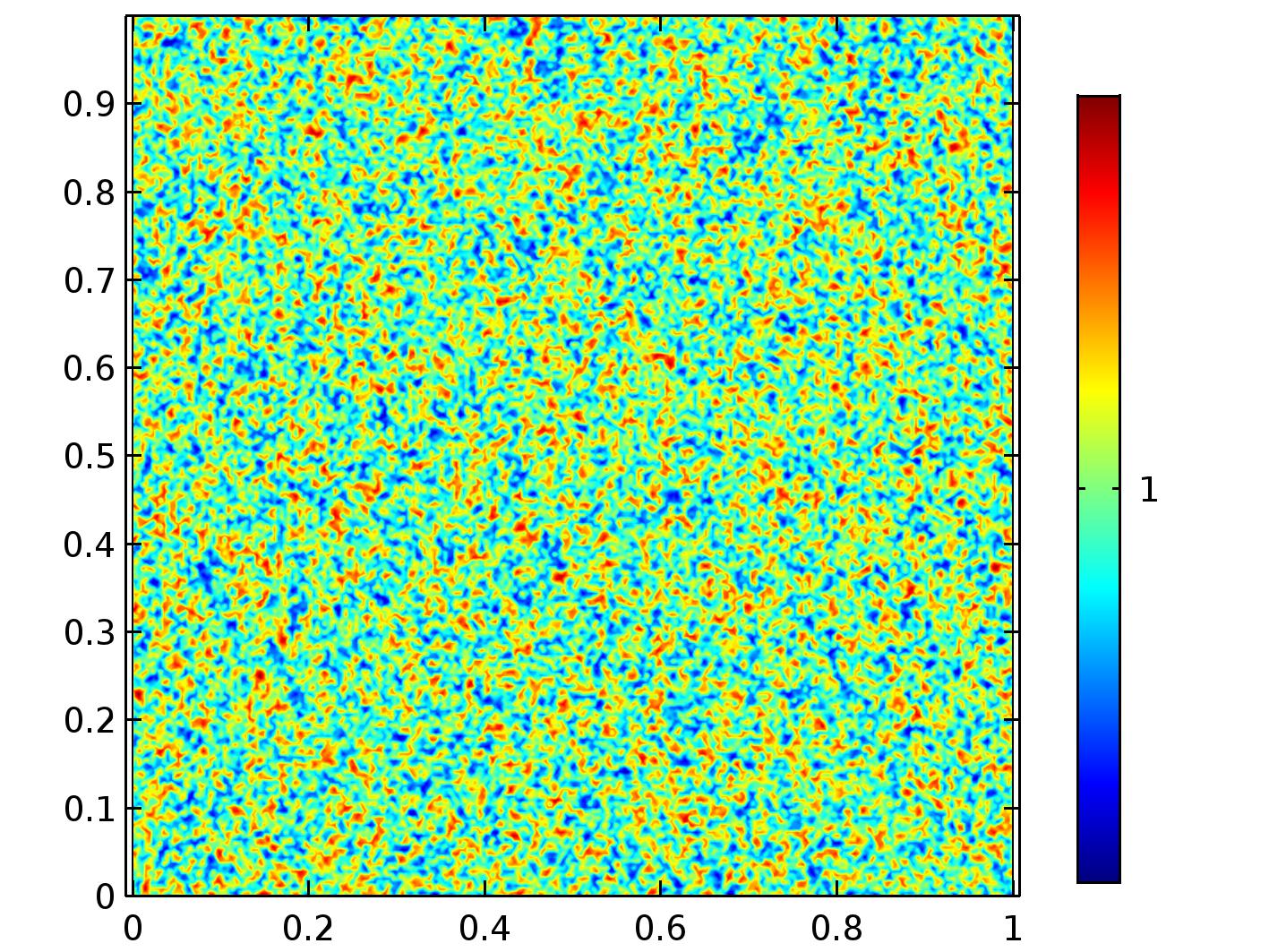} \\
				\includegraphics[width=0.2\linewidth]{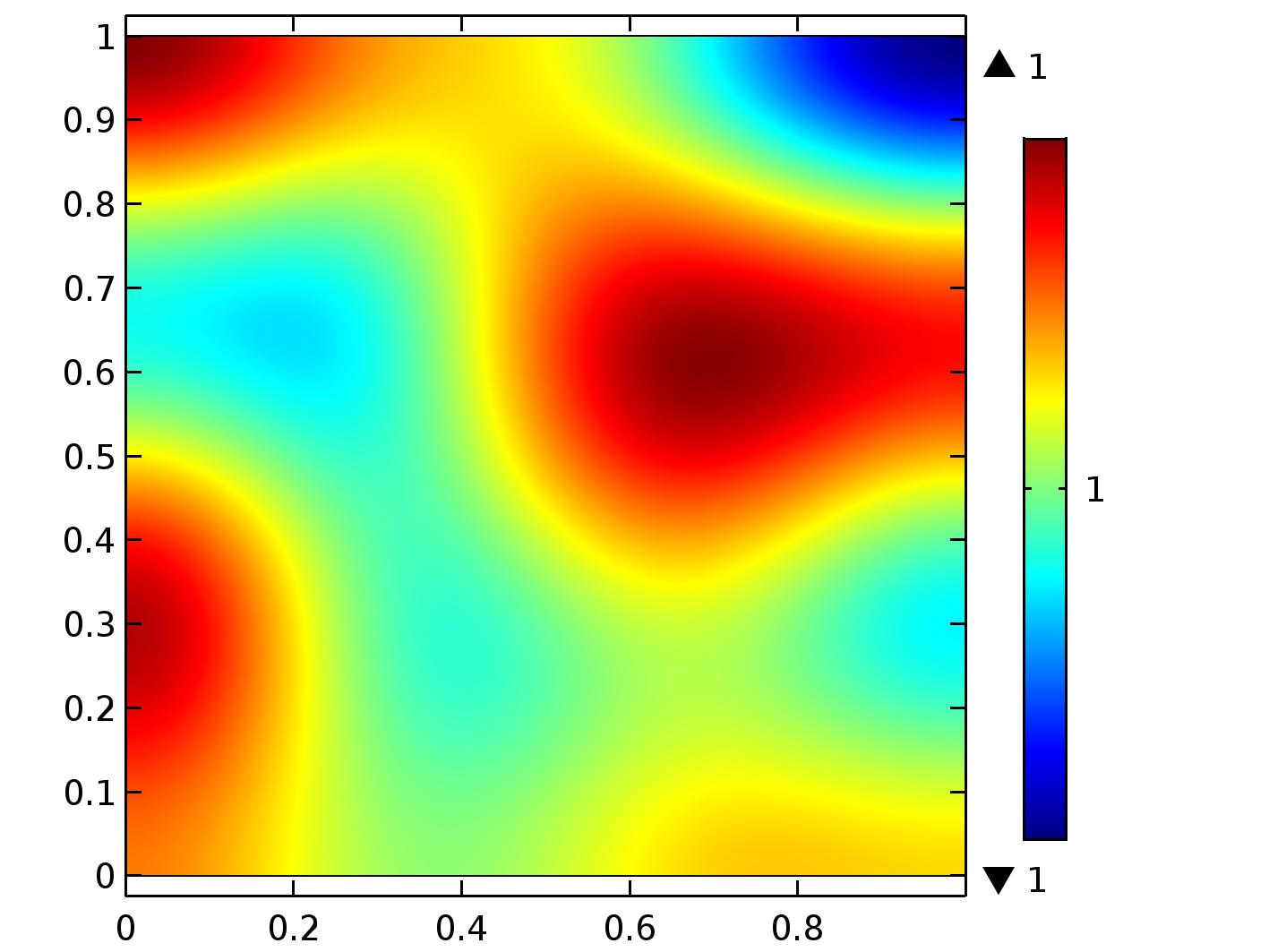} &
        \includegraphics[width=0.2\linewidth]{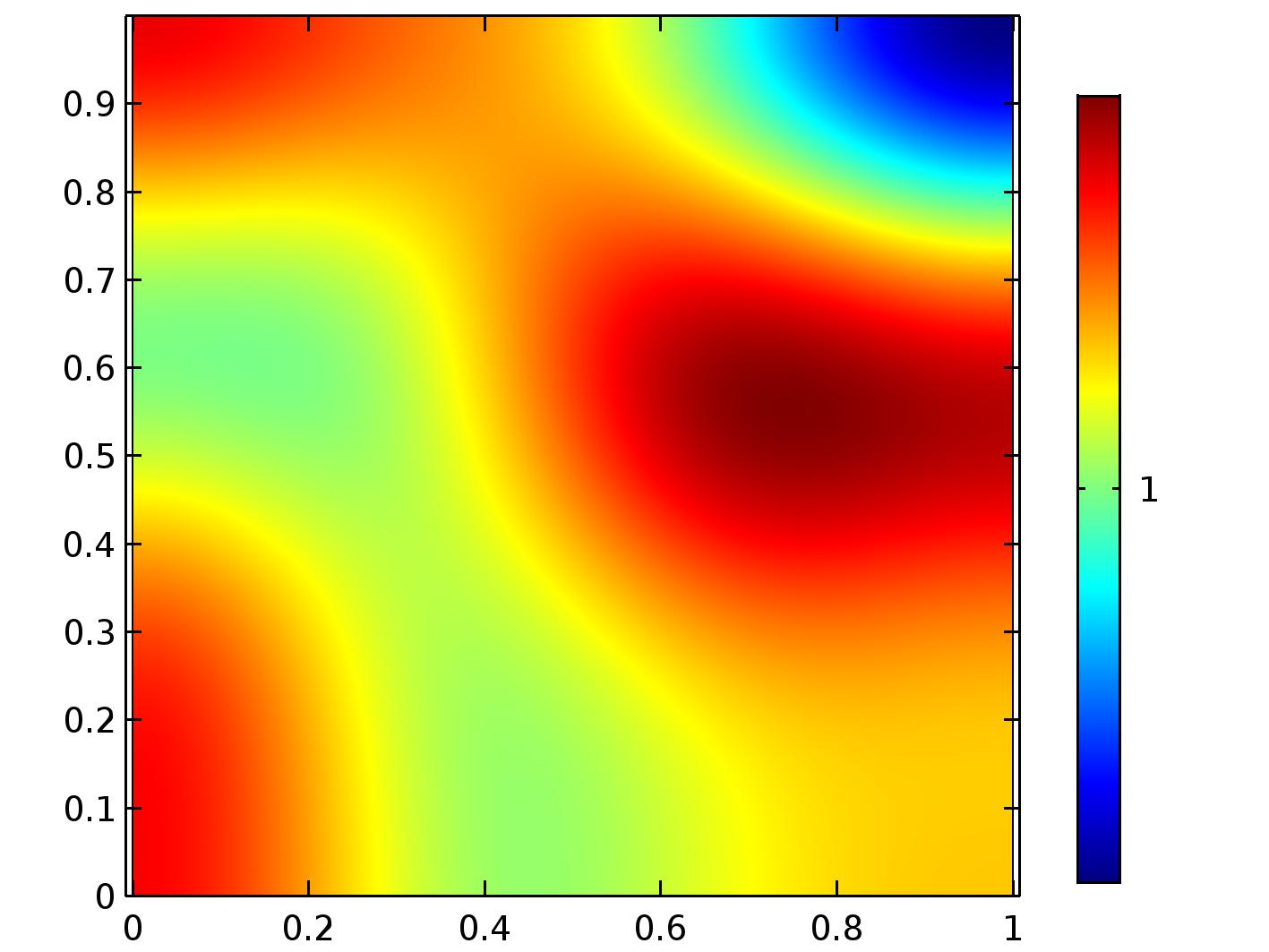} &
        \includegraphics[width=0.2\linewidth]{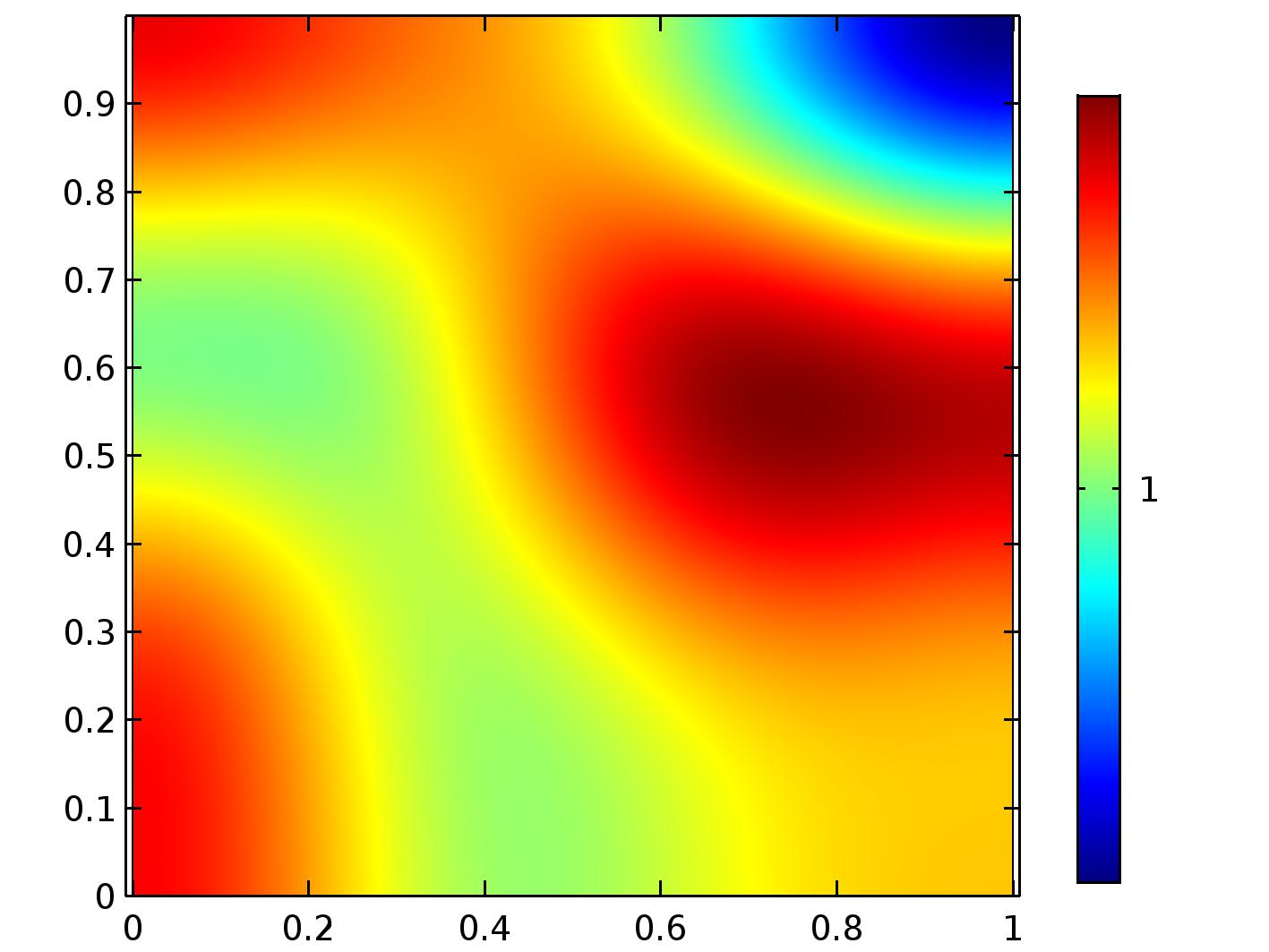} \\
        \includegraphics[width=0.2\linewidth]{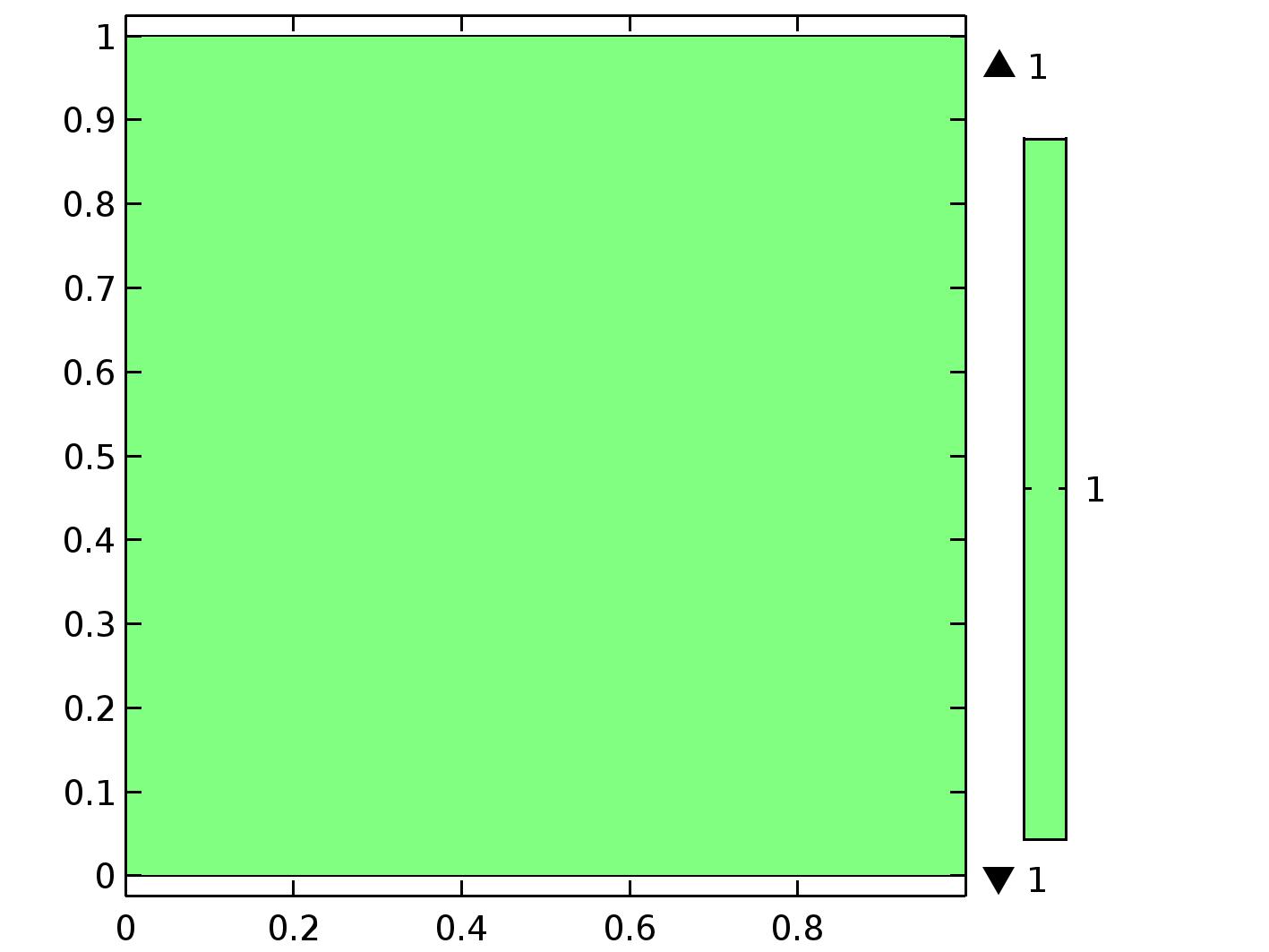} &
        \includegraphics[width=0.2\linewidth]{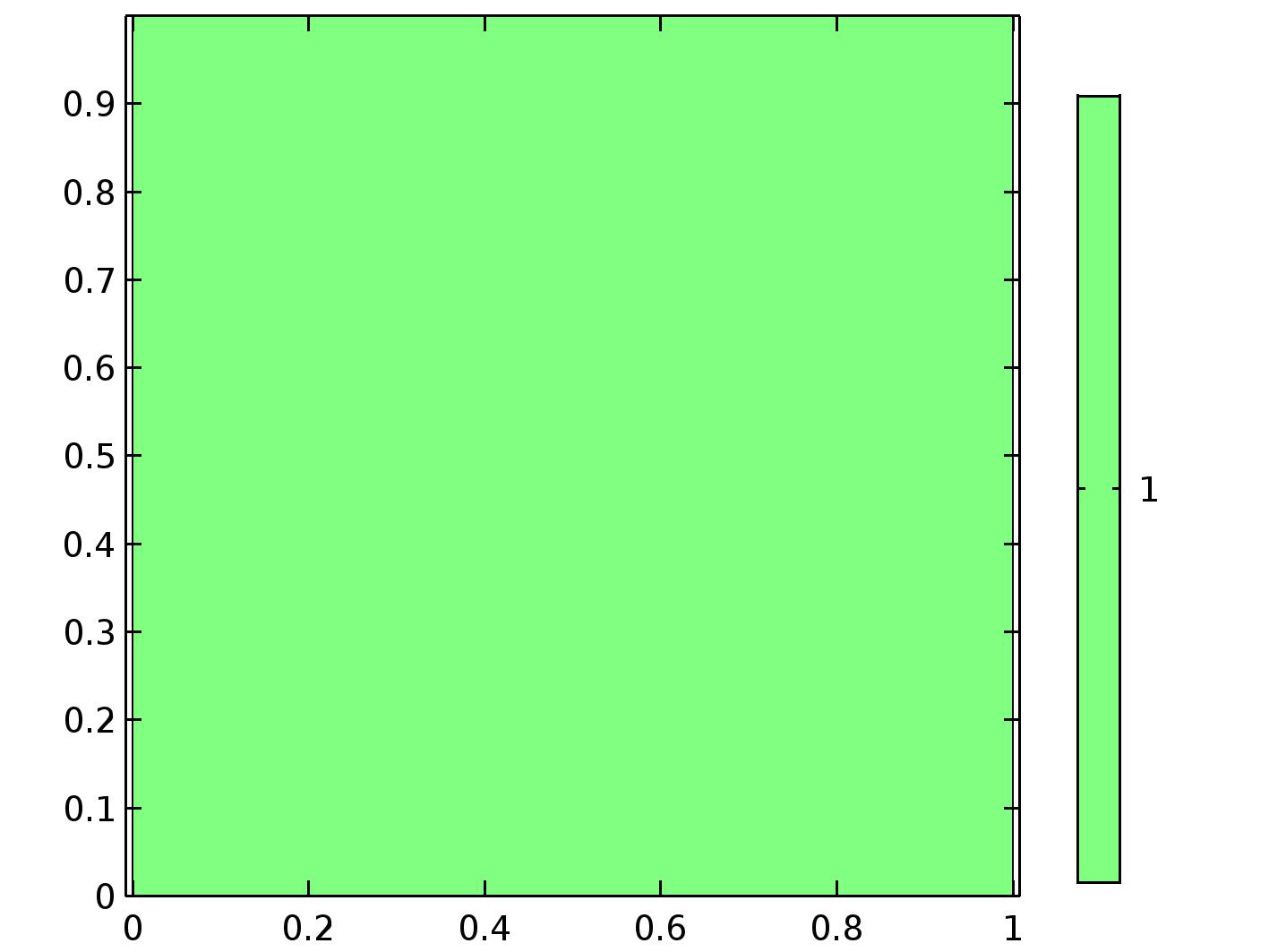} &
        \includegraphics[width=0.2\linewidth]{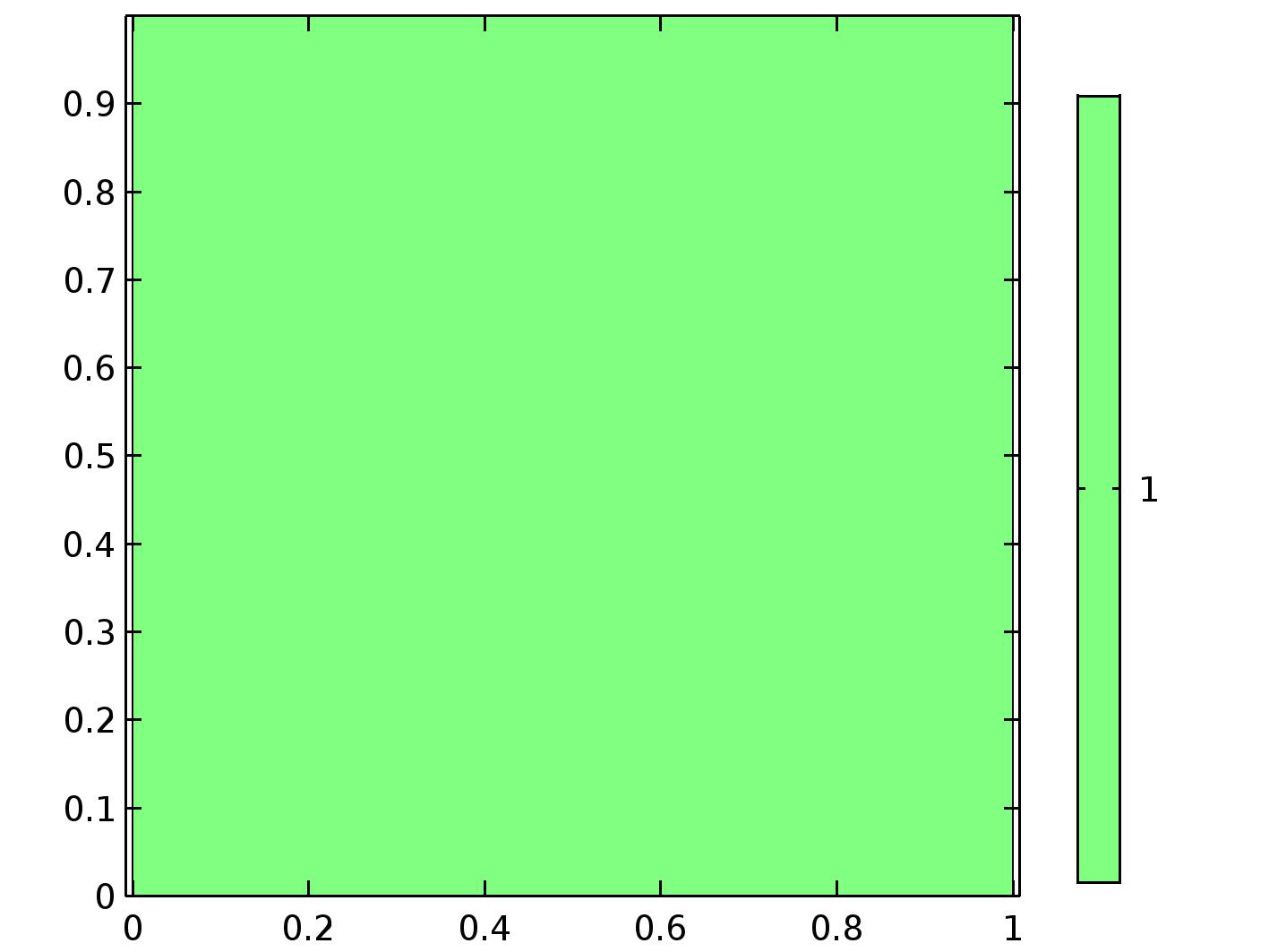} \\
    \end{tabular}
    
    \caption{Time evolution of $CD4+$ Target cells $u$, $CD4+$ Infected cells $v$, and Virions $w$ (column oriented) at $\tau = 0, 0.3, 1, 1.2$.}
    \label{fig:figura1}
\end{figure}
In the second simulation set the chemotactic coefficient was set to $\chi=14 \,mm^4\, cell^{-1}\, day^{-1}$. In this case, a typical pattern formation where the cellular/viral species clearly organized in spots over the spatial domain, emerges (see Figure $2$ and $3$).

\begin{figure}[htbp]
    \centering
    \begin{tabular}{cccccc}
        \includegraphics[width=0.2\linewidth]{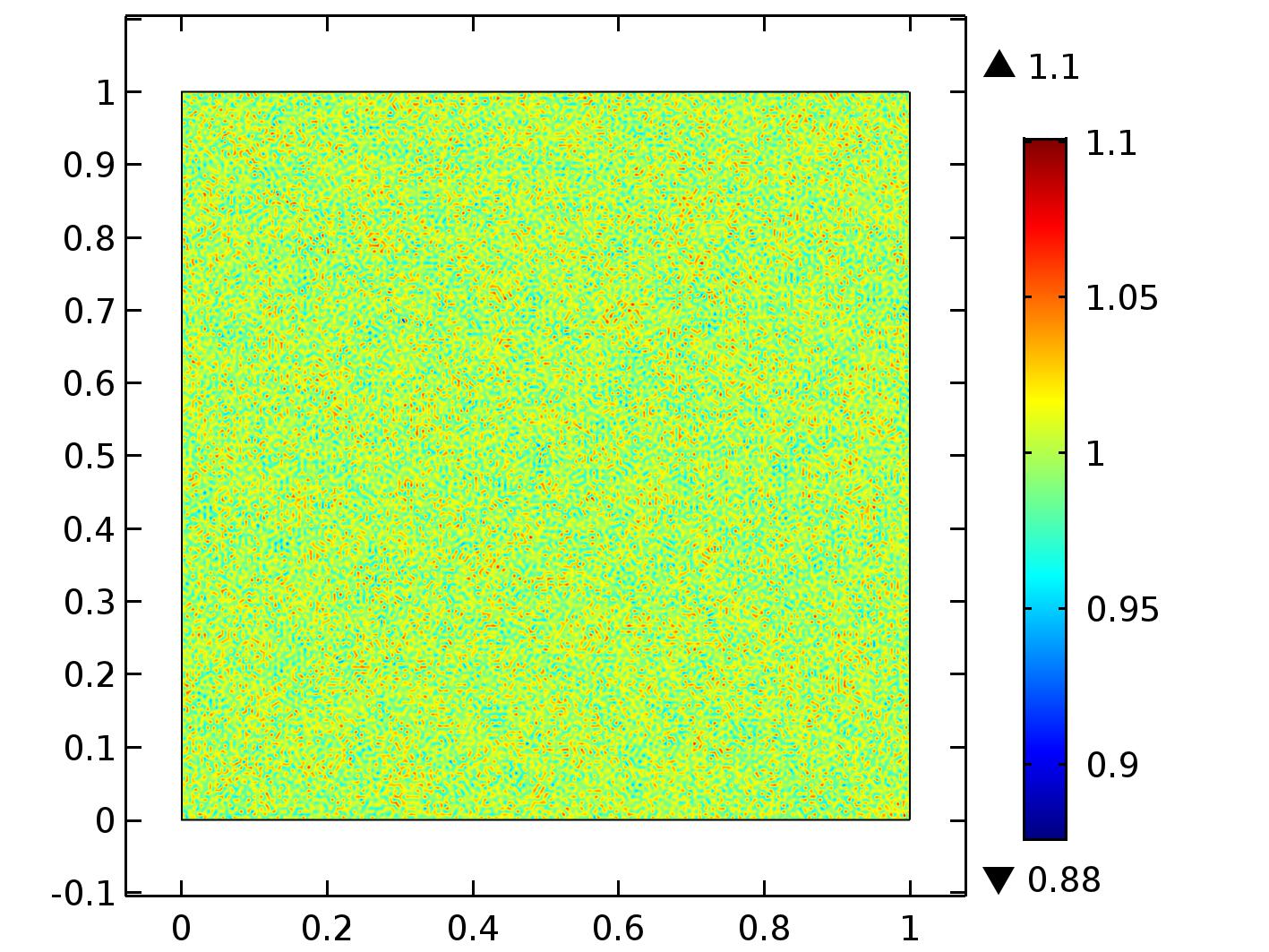} &
				\includegraphics[width=0.2\linewidth]{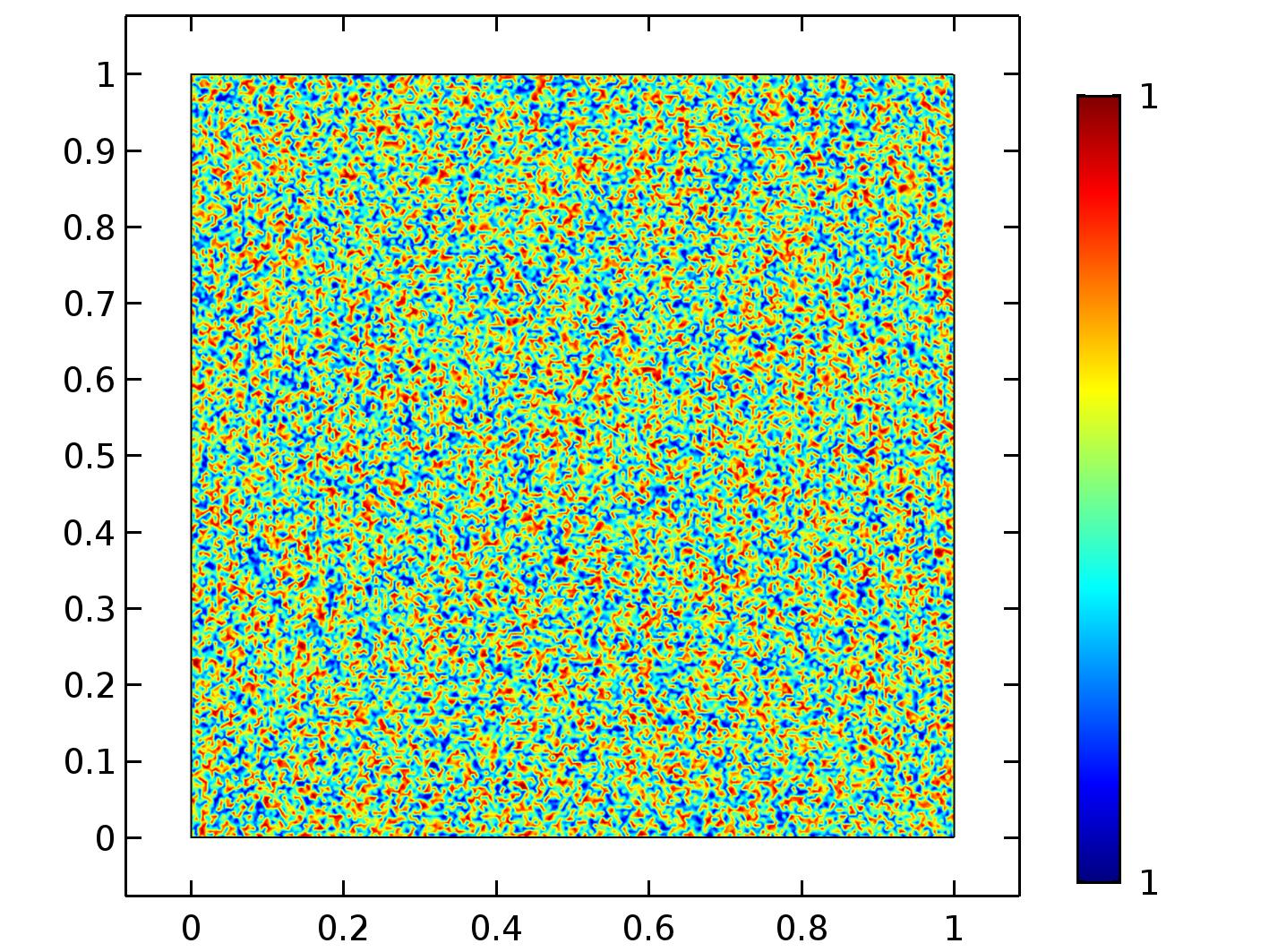} &
        \includegraphics[width=0.2\linewidth]{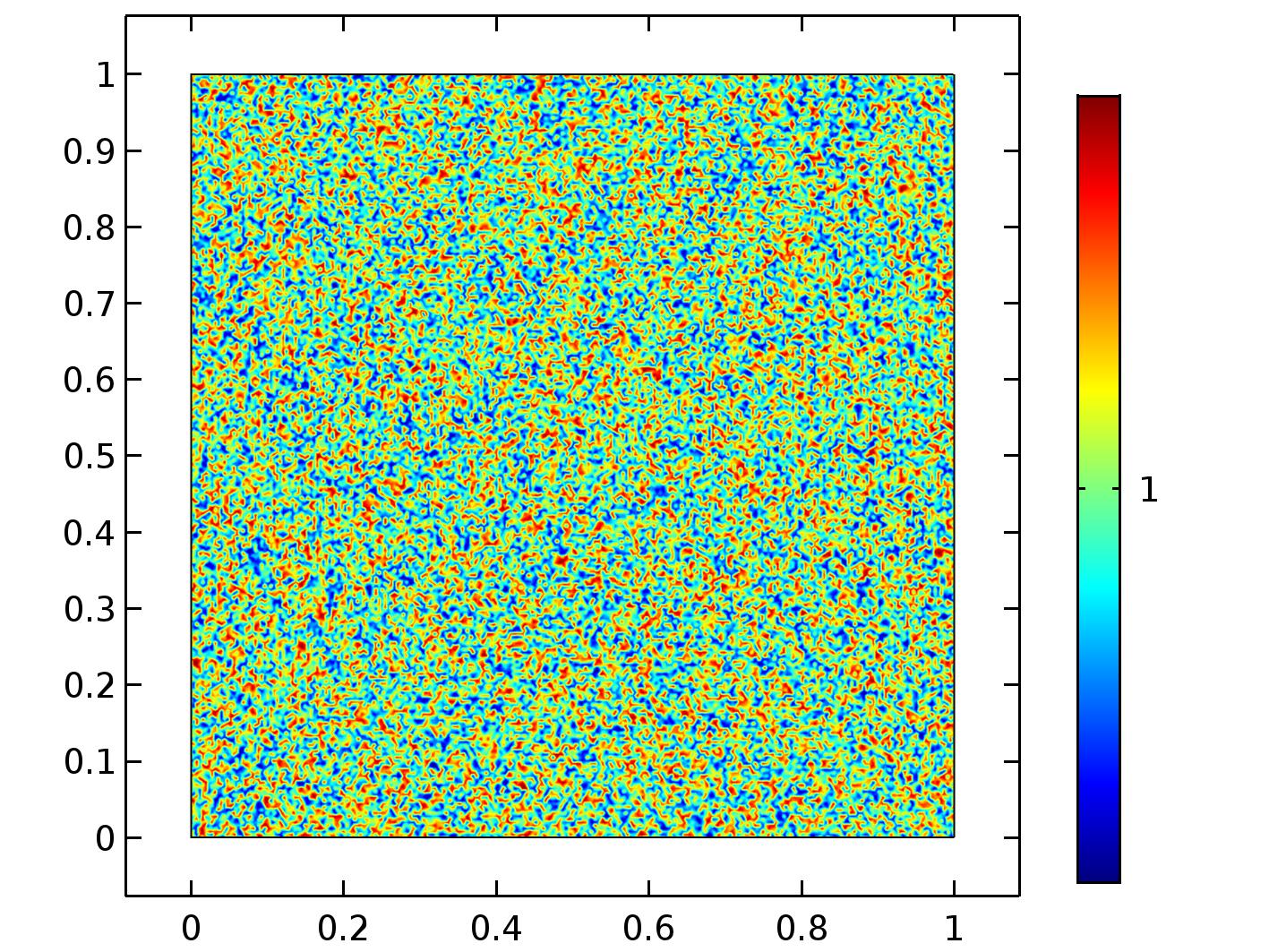} \\
				\includegraphics[width=0.2\linewidth]{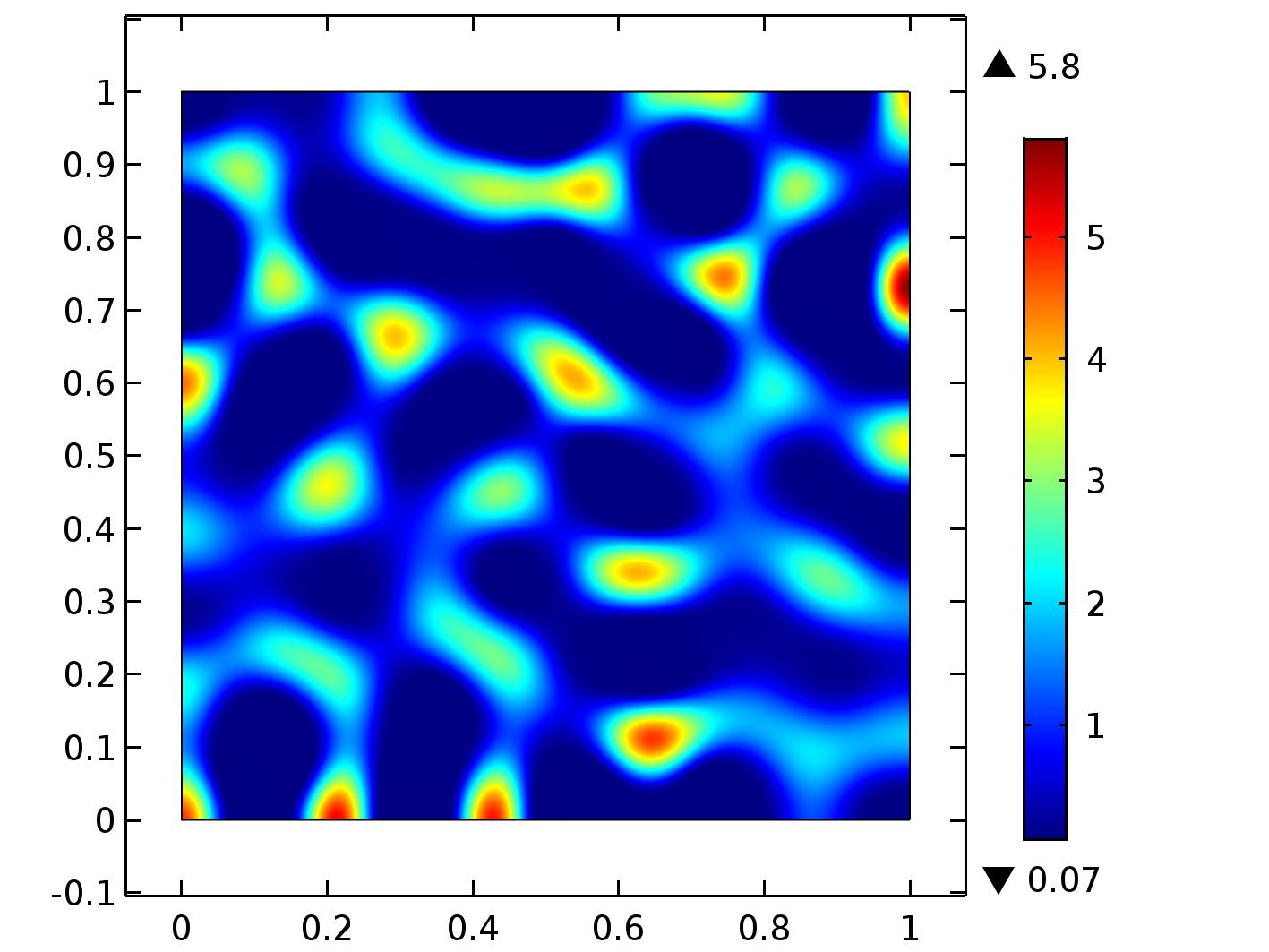} &
        \includegraphics[width=0.2\linewidth]{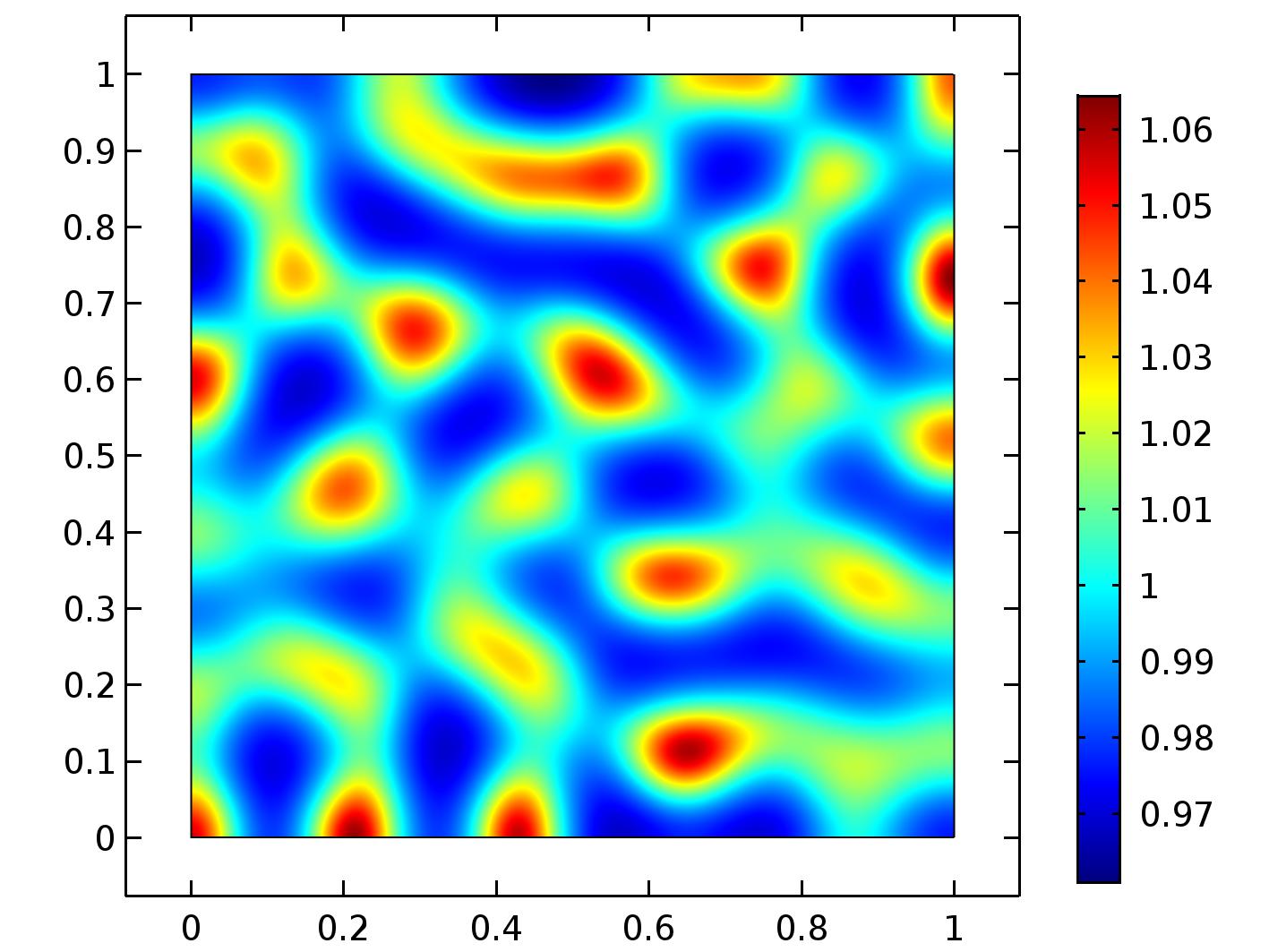} &
				\includegraphics[width=0.2\linewidth]{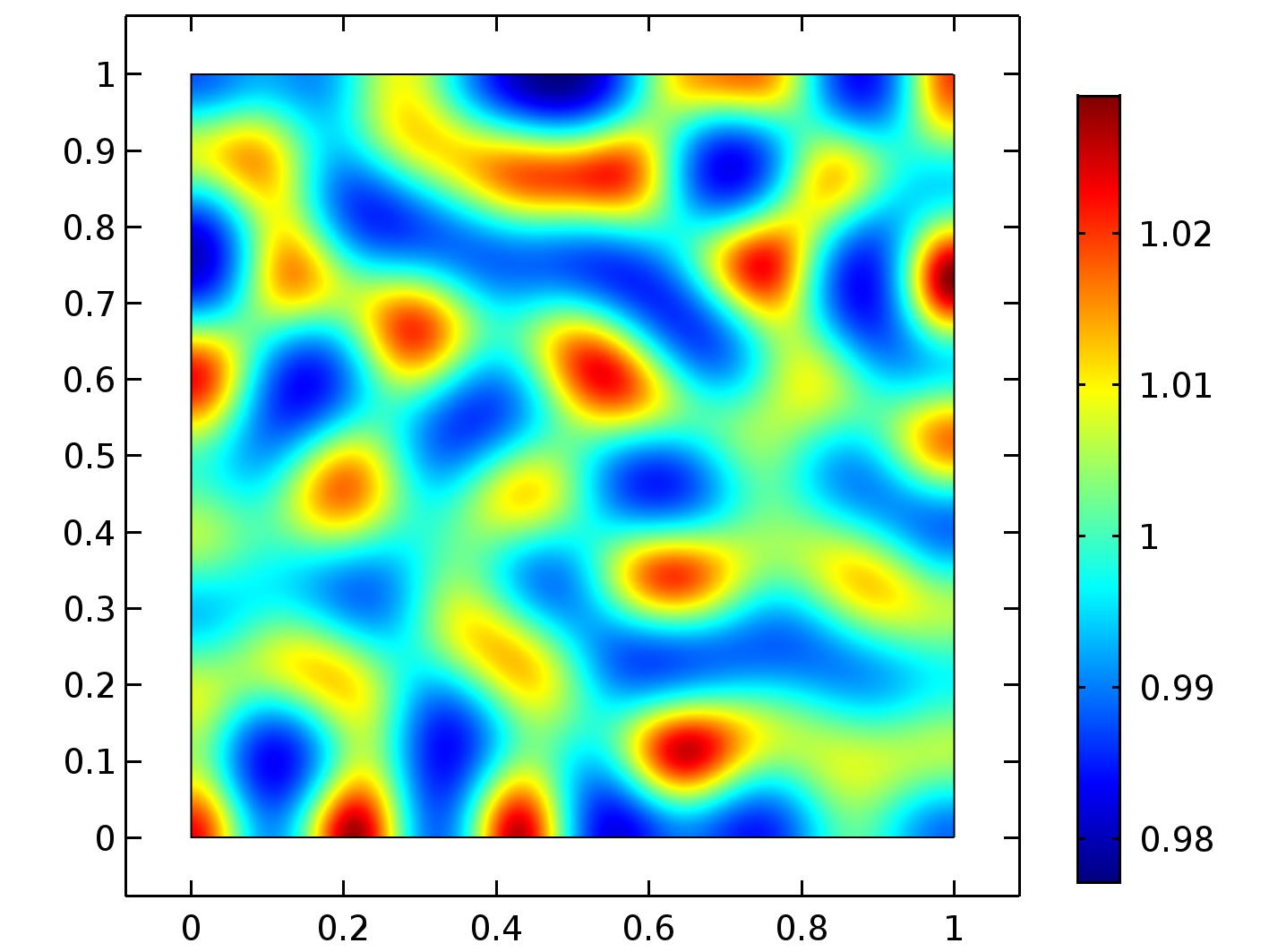} \\
        \includegraphics[width=0.2\linewidth]{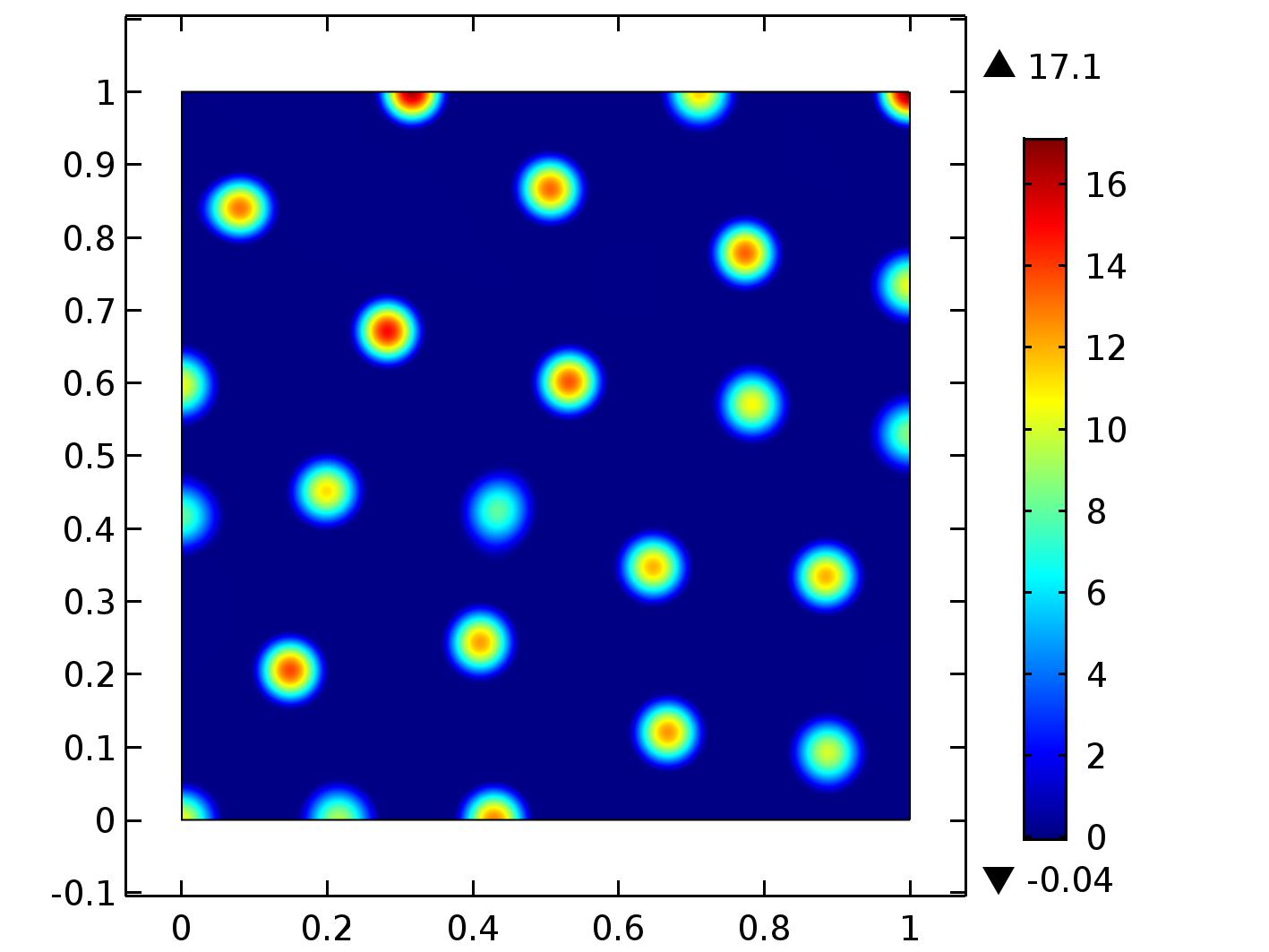} &
        \includegraphics[width=0.2\linewidth]{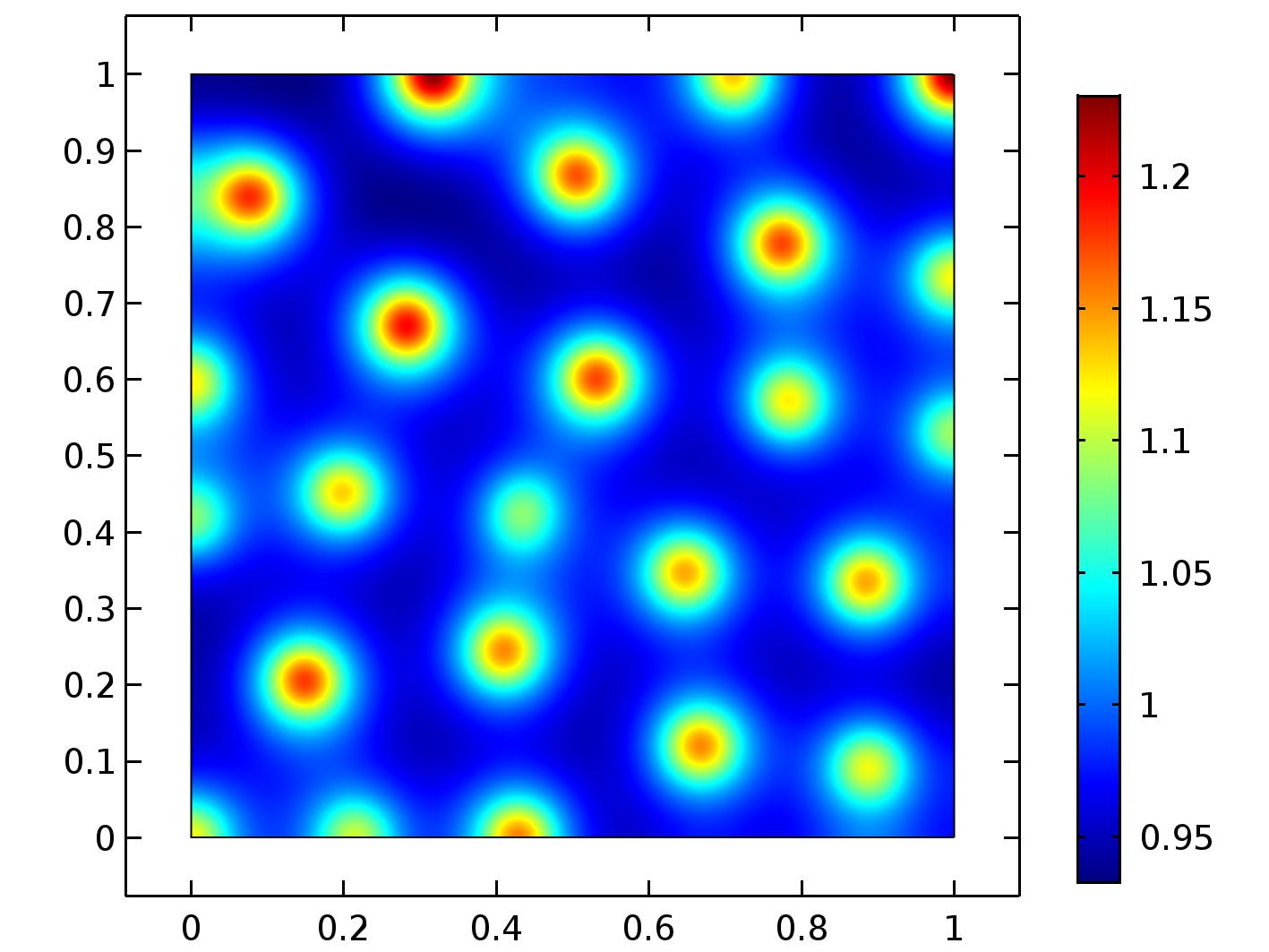} &
        \includegraphics[width=0.2\linewidth]{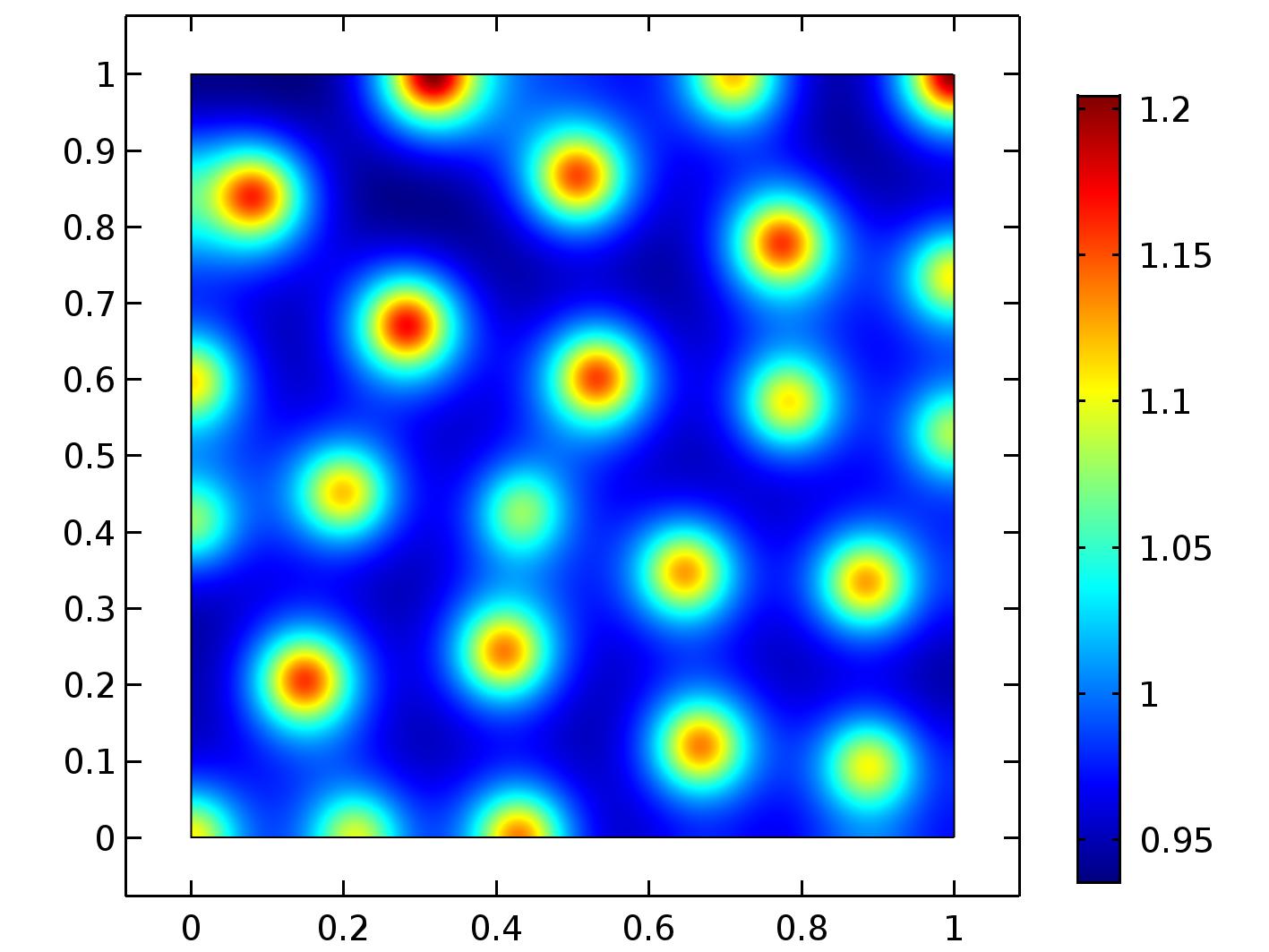} \\
				
    \end{tabular}
    
    \caption{Time evolution of $CD4+$ Target cells $u$, $CD4+$ Infected cells $v$, and Virions $w$ (column oriented) at $\tau = 0, 40, 50, 150$.}
    \label{fig:figura2}
\end{figure}
\noindent
In a 3D plot the evolution of the $CD4+$ Target cells $u$ the accumulation of the investigated species over time in the spatial domain is much more clear, please refer to Figure $3$.

\begin{figure}[htbp]
    \centering
    \begin{tabular}{ccc}
        \includegraphics[width=0.3\linewidth]{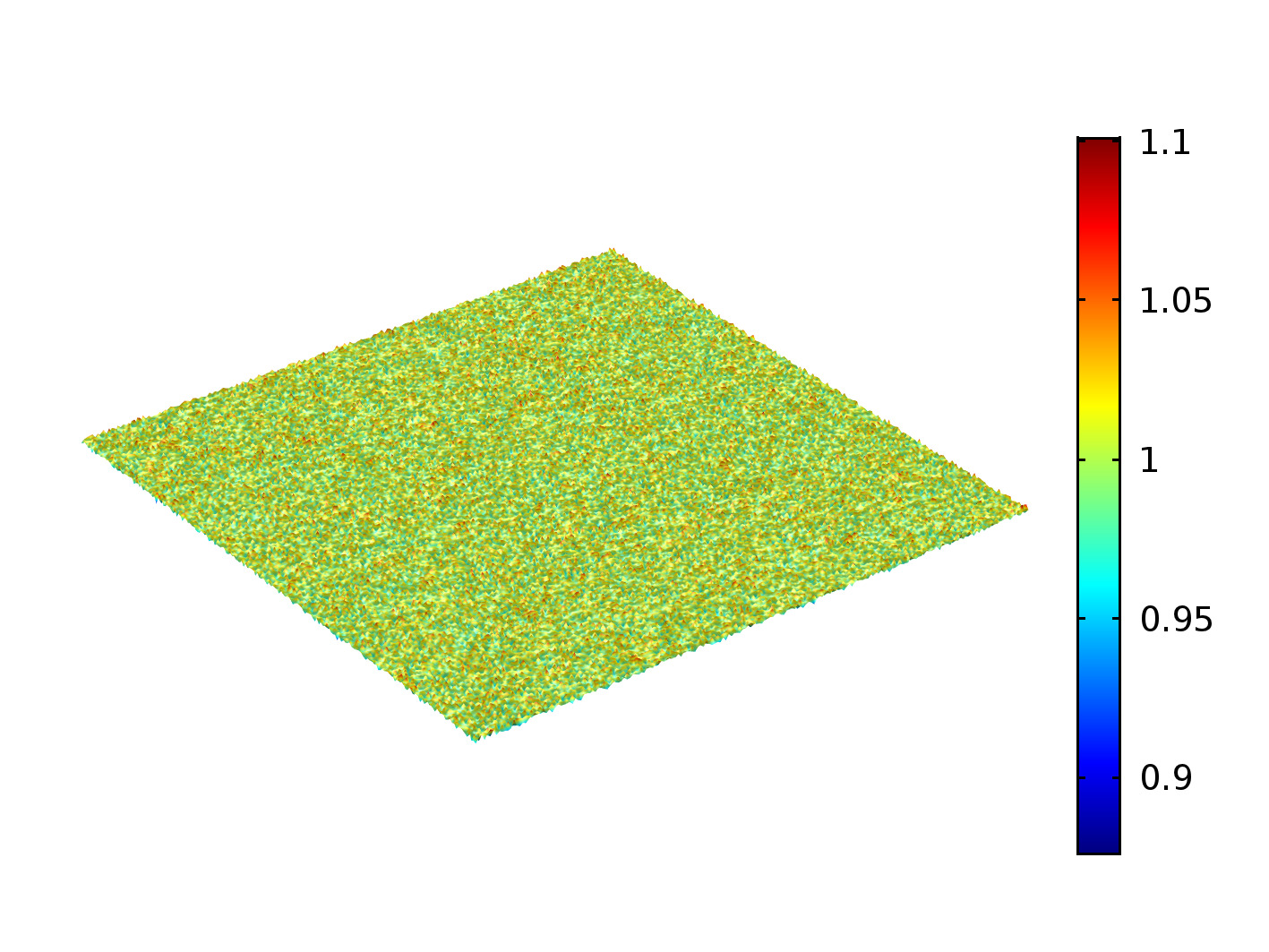} &
        \includegraphics[width=0.3\linewidth]{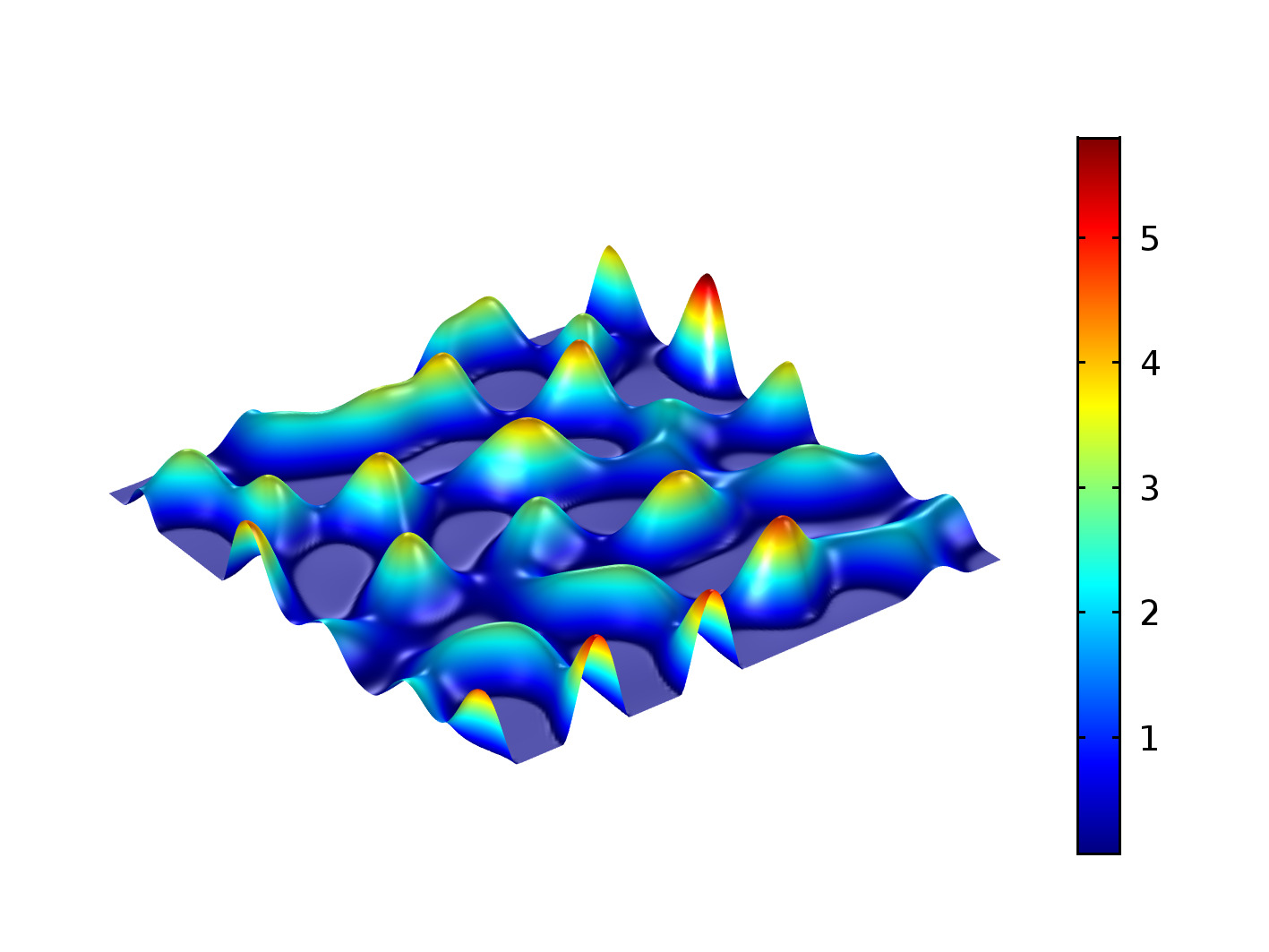} &
        \includegraphics[width=0.3\linewidth]{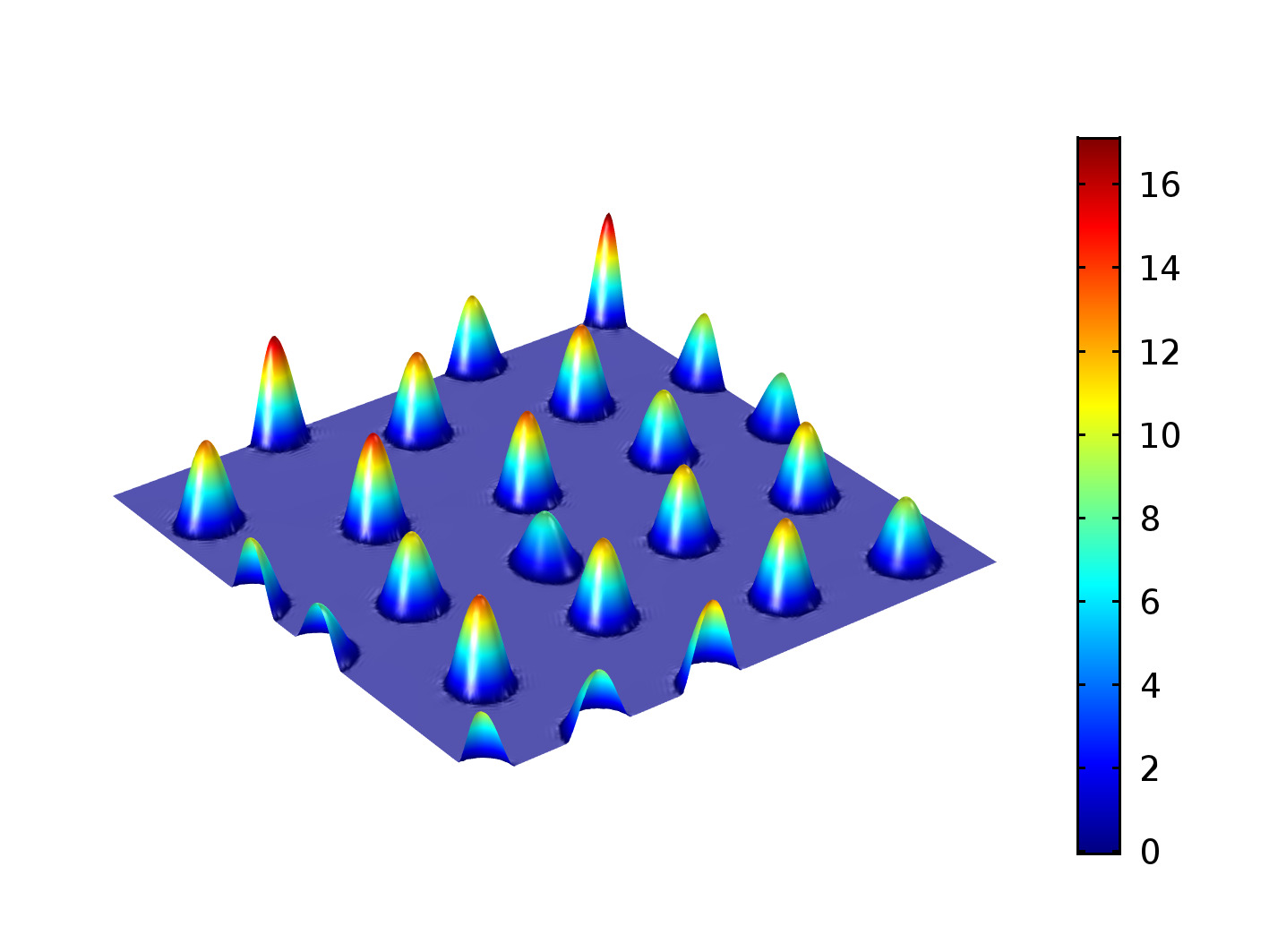} \\
    \end{tabular}
    
    \caption{Time evolution of $CD4+$ Target cells $u$ at $\tau=0, 40, 50, 150$.}
    \label{fig:figura3}
\end{figure}

Finally, another numerical application was performed to investigate the effect of a central infection perturbing the endemic steady state. A Gaussian central perturbation was assumed for $v$, and the diffusive-chemotactic effect leads to the formation of spreading circles in the spatial domain with the definition of a well-defined Turing pattern characterized by a clear radial symmetry, Figure $4$.

\begin{figure}[htbp]
    \centering
    \begin{tabular}{ccc}
        \includegraphics[width=0.3\linewidth]{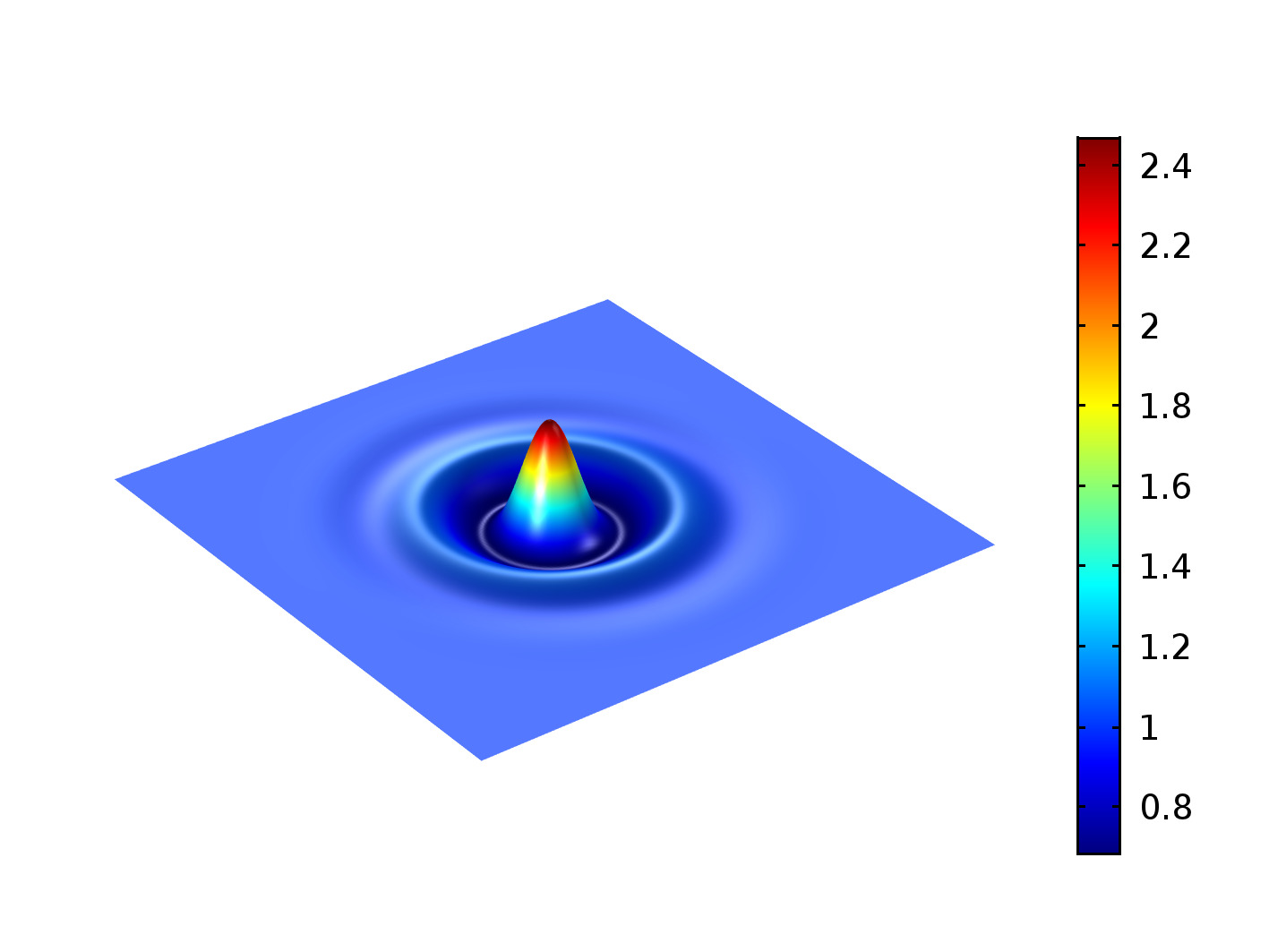} &
        \includegraphics[width=0.3\linewidth]{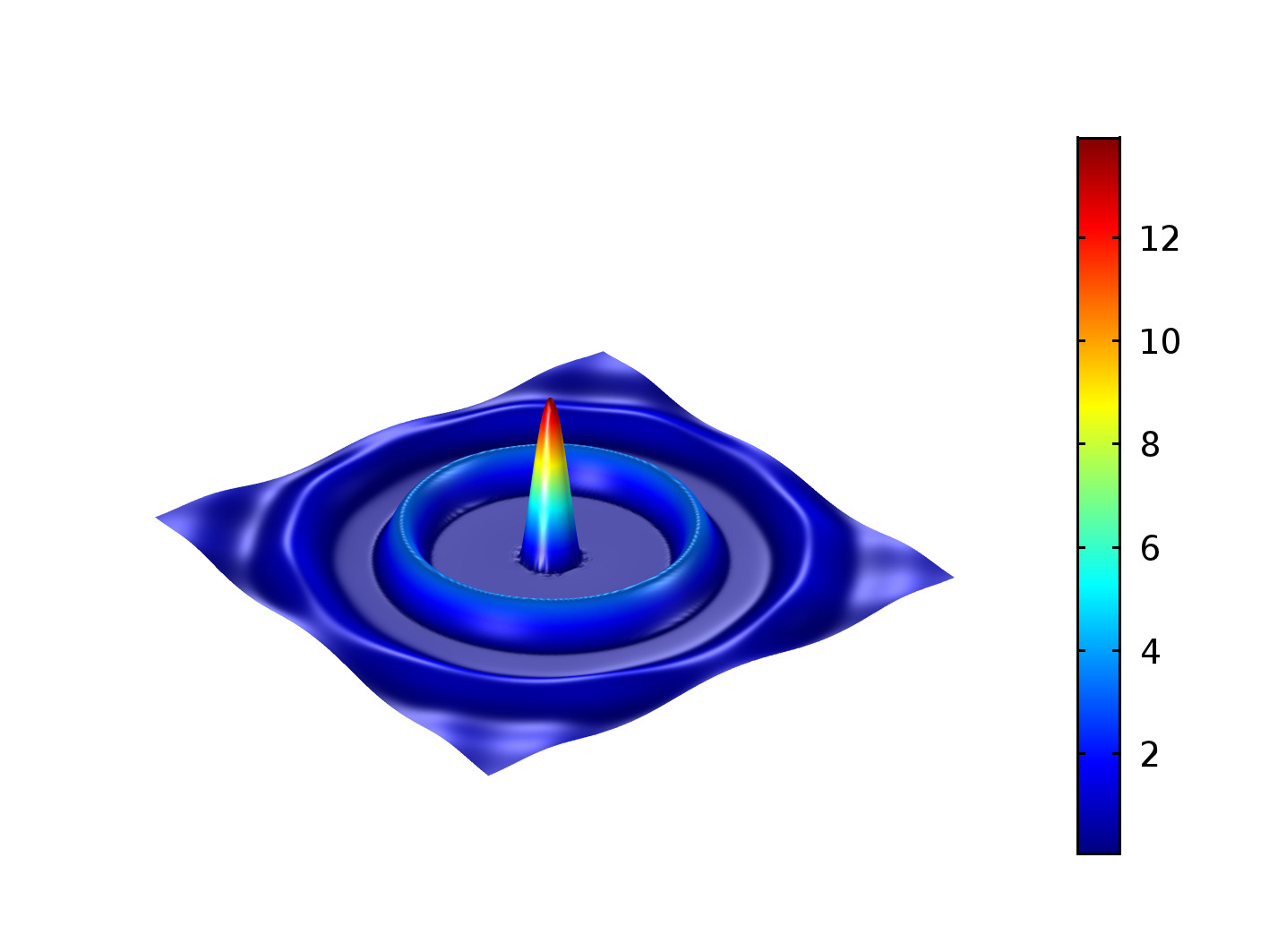} &
        \includegraphics[width=0.3\linewidth]{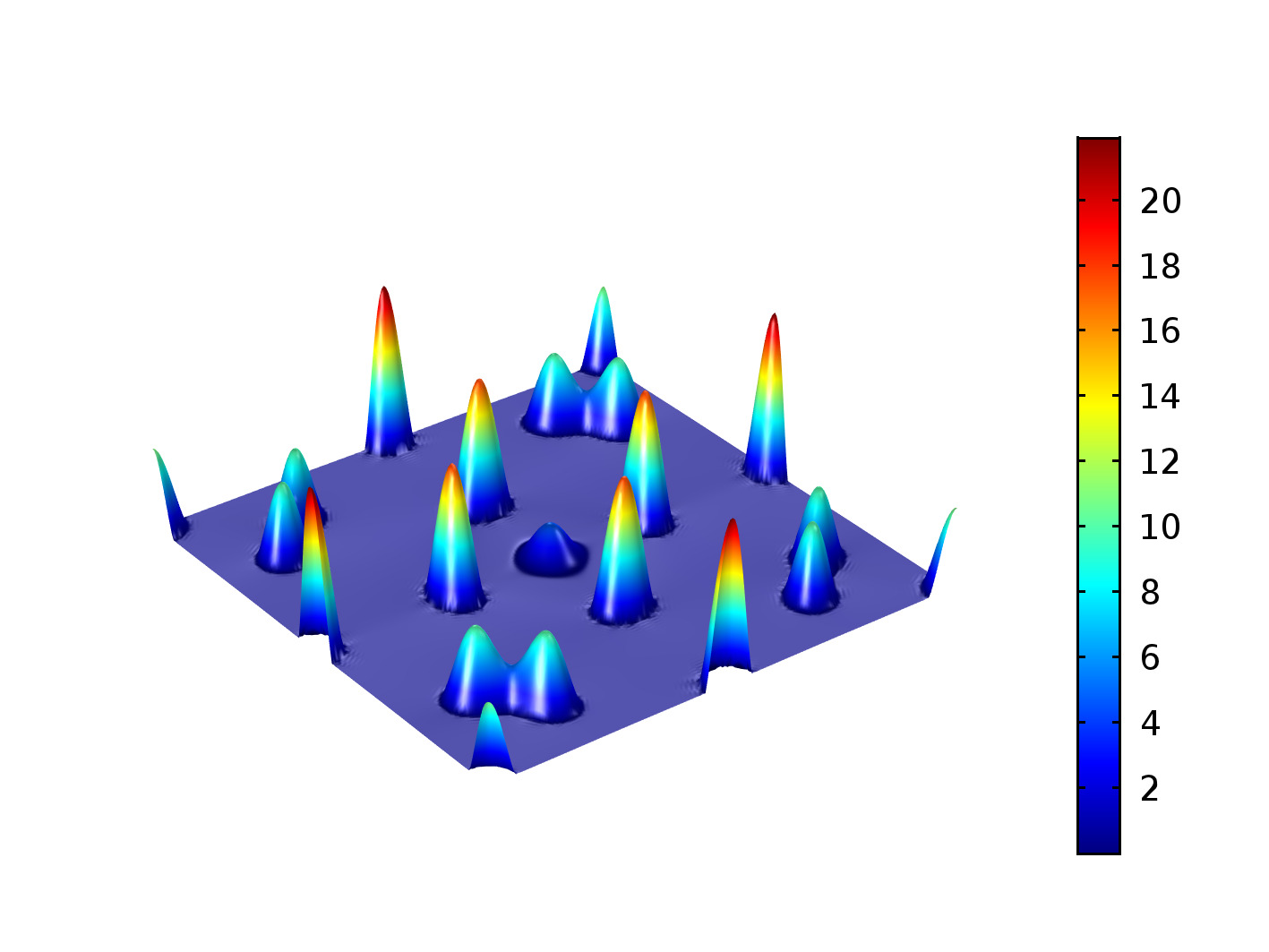} \\
    \end{tabular}
    
    \caption{Time evolution of $CD4+$ Target cells $u$ at $\tau=10, 40, 80, 125$.}
    \label{fig:figura4}
\end{figure}

\section{Preliminaries to the linear instability analysis}

The preliminary numerical investigation shows that the model exhibits spatial patterns when $\chi$ overcomes a certain critical value. In the next sections, we aim to find this threshold in closed form.

Introducing $U,V,W$ the perturbation fields to the generic equilibrium $(u^*,v^*,w^*)$, the linearized no-dimensional perturbation system is
\begin{equation}\label{pertsys}
\dfrac{\partial\mathbf U}{\partial t}=\mathcal L\mathbf U+\mathcal D\Delta \mathbf U,\end{equation}
with 
$$
\mathbf U=(U,V,W)^T,\,\,\mathcal L=\left(\begin{array}{ccc}
                  a_{11}&0&a_{13}\\
                  a_{21}&a_{22}&a_{23}\\
                  0&a_{32}&a_{33}
                 \end{array}
\right),\quad \mathcal D=\left(\begin{array}{ccc}
                                d_1&-dI_cu^*&0\\
                                0&d_2&0\\
                                0&0&d_3
                               \end{array}
\right)
$$
and
\begin{equation}\label{coefflinsys}
\begin{array}{l}
 a_{11}=-1+\delta\left(1-\dfrac{2u^*}{h}\right)-\eta I_cw^*,\,\,
 a_{13}=-\eta I_c u^*,\,\, a_{21}=\mu_1 w^*,\\a_{22}=-\mu_1,\,\,
 a_{23}=\mu_1 u^*,\,\,a_{32}=\mu_2,\,\,a_{33}=-\mu_2.
\end{array}\end{equation}

To look for Turing instability, we have preliminarily explore the stability of the biological meaningful equilibria in the homogeneous case.

\subsection{Stability analysis of disease-free equilibrium in the homogeneous case}

Linearizing about the homogeneous disease-free equilibrium, one obtains 
\begin{equation}\label{linsysdfehom}
\dfrac{\partial\mathbf U}{\partial t}=\mathcal L_0\mathbf U,\end{equation}
with $\mathcal L_0$ is given by $\mathcal L$ evaluated in $E_0=(\bar u, 0,0)$. The $\mathcal L_0-$eigenvalues are
$$
\lambda_1= -1+\delta\left(1-\dfrac{2\bar u}{h}\right)=
-\dfrac{\sqrt{h^2(\delta-1)^2+4h\delta\xi}}{h}<0$$
and the solutions of the equation
\begin{equation}\label{eqcardfehom}
\lambda^2+(\mu_1+\mu_2)\lambda+\mu_1\mu_2(1-\bar u)=0.\end{equation}
The two real roots of (\ref{eqcardfehom}) are negative if and only if $\bar u<1$. In view of 
$$
\begin{array}{l}
 1-\bar u=\dfrac{2\delta-\left[h(\delta-1)+\sqrt{h^2(\delta-1)^2+4h\delta\xi}\right]}{2\delta},
\end{array}
$$
it turns out that $E_0$ is homogeneously, linearly stable if and only if $R_0<1$.

\subsection{Stability analysis of the endemic equilibrium in the homogeneous case}

Linearizing about the homogeneous endemic equilibrium (existing only if (\ref{exist}) holds), one obtains 
\begin{equation}\label{linsyseehom}
\dfrac{\partial\mathbf U}{\partial t}=\mathcal L_c\mathbf U,\end{equation}
with $\mathcal L_c$ is given by $\mathcal L$ evaluated in $E_c=(1, 1, 1)$, i.e.,

$\mathcal L_c=\left(\begin{array}{ccc}
                    -\dfrac{\delta}{h}-\xi&0&-\eta I_c\\
                    \mu_1&-\mu_1&\mu_1\\
                    0&\mu_2&-\mu_2
                   \end{array}
\right).$
\noindent
Setting $\texttt I_1, \texttt I_2, \texttt I_3 $ the $\mathcal L_c-$ principal invariants, i.e.
\begin{equation}\label{invahomee}
\begin{array}{l}
 \texttt I_1=\mbox{trace }\mathcal L_c=-\left(\dfrac{\delta}{h}+\xi+\mu_1+\mu_2\right),\,
 \texttt I_2=(\mu_1+\mu_2)\left(\dfrac{\delta}{h}+\xi\right),\\
 \texttt I_3=\det \mathcal L_c=-\eta I_c\mu_1\mu_2,
\end{array}
\end{equation}
the Routh-Hurwitz conditions, \emph{necessary and sufficient} to guarantee that all the $\mathcal L_c$-eigenvalues have negative real part, are \cite{Merkin}
\begin{equation}\label{RHeehom}
\texttt I_1<0,\quad \texttt I_3<0,\quad \texttt I_1\texttt I_2-\texttt I_3<0,\end{equation}
which necessarily require that $\texttt I_2>0$. It is easy to check that the first two conditions in (\ref{RHeehom}) are verified and, in view of (\ref{12*})
$$
\begin{array}{l}
 \texttt I_1\texttt I_2\!-\!\texttt I_3\!=\!-(\mu_1\!+\!\mu_2)\left(\xi\!+\!\dfrac{\delta}{h}\right)^2\!-\!(\mu^2_1\!+\!\mu^2_2)\left(\xi\!+\!\dfrac{\delta}{h}\right)
\! -\!\mu_1\mu_2\left(\xi\!+\!1\!-\!\delta\!+\!\dfrac{3\delta }{h}\right)\!<\!0.
\end{array}
$$
\noindent
Then the endemic equilibrium is \emph{always homogeneously, linearly stable} when it exists.

Collecting the results obtained in subsections 5.1 and 5.2, we proved: \emph{in the absence of spatial variations, the disease-free equilibrium is linearly stable if and only if $R_0<1$. When $R_0>1$, the disease-free equilibrium becomes unstable and there exists a unique, linearly stable, endemic equilibrium}.

\section{Linear stability analysis in the heterogeneous case}

In this section we look for conditions guaranteeing that an equilibrium -- stable in the homogeneous case -- becomes unstable in presence of self-diffusion and chemotaxis (Turing instability).

\subsection{Stability analysis of the disease-free equilibrium in the heterogeneous case}

System (\ref{pertsys}) evaluated in $E_0$ becomes
\begin{equation}\label{linsysdfe}
\dfrac{\partial\mathbf U}{\partial t}=\mathcal L_0\mathbf U+\mathcal D_0\Delta \mathbf U,\qquad \mathcal D_0=\left(\begin{array}{ccc}
                                d_1&-dI_c\bar u&0\\
                                0&d_2&0\\
                                0&0&d_3
                               \end{array}
\right).\end{equation}
The dispersion relation governing the eigenvalues $\lambda$ in terms of the wave number $k$ is
\begin{equation}\label{eqcardfe}
\lambda^3-T^0_k(k^2)\lambda^2+\texttt I^0_k(k^2)\lambda-h^0(k^2)=0,\end{equation}
where
$$
\begin{array}{ll}
 T^0_k(k^2)&=-\left(\dfrac{\sqrt{h^2(\delta-1)^2+4h\delta\xi}}{h}+\mu_1+\mu_2\right)-(d_1+d_2+d_3)k^2,\\
 \texttt I^0_k(k^2)&=(d_1d_2+d_1d_3+d_2d_3)k^4\\
 &+\left[d_1(\mu_1+\mu_2)+d_2\mu_2+d_3\mu_1+(d_2+d_3)\dfrac{\sqrt{h^2(\delta-1)^2+4h\delta\xi}}{h}\right]\\
 &+(\mu_1+\mu_2)\dfrac{\sqrt{h^2(\delta-1)^2+4h\delta\xi}}{h}+\mu_1\mu_2(1-\bar u)\\
 h^0(k^2)&=-d_1d_2d_3k^6-\left[d_2d_3\dfrac{\sqrt{h^2(\delta-1)^2+4h\delta\xi}}{h}+d_1(d_2\mu_2+d_3\mu_1)\right]k^4\\
 &-\left[d_1\mu_1\mu_2(1-\bar u)+\dfrac{\sqrt{h^2(\delta-1)^2+4h\delta\xi}}{h}(d_2\mu_2+d_3\mu_1)\right]k^2\\
 &-\dfrac{\sqrt{h^2(\delta-1)^2+4h\delta\xi}}{h}\mu_1\mu_2(1-\bar u).
\end{array}
$$
When $R_0<1$, it follows that 
\begin{equation}
T^0(k^2)<0,\quad h^0(k^2)<0,\quad T^0(k^2)\texttt I^0(k^2)-h^0(k^2)<0,\forall k^2\in\mathbb R^+.
\end{equation}
Then: \emph{if and only if $R_0<1$, the disease-free equilibrium is linearly stable both in the absence and in the presence of diffusion}.

\subsection{Stability analysis for the endemic equilibrium and Turing instability}

System (\ref{pertsys}) evaluated in $E_c=(1,1,1)$ becomes
\begin{equation}\label{linsysee}
\dfrac{\partial\mathbf U}{\partial t}=\mathcal L_c\mathbf U+\mathcal D_c\Delta \mathbf U,\qquad \mathcal D_c=\left(\begin{array}{ccc}
                                d_1&-dI_c&0\\
                                0&d_2&0\\
                                0&0&d_3
                               \end{array}
\right).\end{equation}
The dispersion relation governing the eigenvalues $\lambda$ in terms of the wave number $k$ is
\begin{equation}\label{eqcaree}
\lambda^3-T_k(k^2)\lambda^2+\texttt I^c_k(k^2)\lambda-h(k^2)=0,\end{equation}
where
$$
\begin{array}{ll}
 T_k(k^2)&=\texttt I_1-(d_1+d_2+d_3)k^2,\\
 \texttt I^c_k(k^2)&=(d_1d_2+d_1d_3+d_2d_3)k^4+\texttt I_2\\
 &+\left[(\mu_1+\mu_2)d_1+\mu_2d_2+\mu_1d_3+(d_2+d_3)\left(\dfrac{\delta}{h}+\xi\right)-\mu_1 d I_c\right]k^2\\
h(k^2)&=-d_1d_2d_3k^6-\left[\left(\dfrac{\delta}{h}+\xi\right)d_2d_3+\mu_1d_1d_3+\mu_2d_1d_2-\mu_1 d d_3 I_c\right]k^4\\
&-\left[\left(\dfrac{\delta}{h}+\xi\right)(d_2\mu_2+d_3\mu_1)-\mu_1\mu_2 dI_c\right]k^2+\texttt I_3
\end{array}
$$
with $\texttt I_1, \texttt I_2, \texttt I_3$ given by (\ref{invahomee}).
The endemic equilibrium is linearly stable if and only if
\begin{equation}\label{RH}
T_k(k^2)<0,\quad h(k^2)<0,\quad T_k(k^2)\texttt I^c_k(k^2)-h(k^2)<0,\quad \forall k^2\in\mathbb R^+.\end{equation}
If one of (\ref{RH}) is reversed for at least one $\bar k\in\mathbb R^+$, then Turing instability occurs. Since (\ref{RH})$_1$ is always verified, \emph{necessary} conditions for the occurrence of diffusion-driven instability can be obtained by requiring that, for at least one $\bar k\in\mathbb R^+$, at least one of the following conditions is verified
\begin{equation}\label{ct}
\texttt I^c_k(\bar k)<0,\quad h(\bar k)>0,\quad 
T_k(\bar k)\texttt I^c_k(\bar k)-h(\bar k)>0.\end{equation}
Setting
$$\begin{array}{l}
\delta_1=d_1d_2+d_1d_3+d_2d_3,\,\,
\delta_2=\tilde\delta-\mu_1 d I_c,\,\, \delta_3=\texttt I_2,\\
\tilde\delta=(\mu_1+\mu_2)d_1+\mu_2d_2+\mu_1d_3+(d_2+d_3)(\xi+\delta/h),
  \end{array}
$$
one has that 
$\texttt I^c_k(k^2)=\delta_1 k^4+\delta_2 k^2+ \delta_3$.
Since $\delta_1>0, \delta_3>0$, a \emph{necessary} condition guaranteeing that $\texttt I^c_k$ can assume some negative values is $\delta_2<0$, i.e.
\begin{equation}\label{cnec1}
d>\dfrac{\tilde\delta}{\mu_1 I_c}
\end{equation}
or, recalling that $\eta I_c=\xi-1+\delta\left(1-\dfrac{1}{h}\right)$,
$
d>\dfrac{\eta}{\mu_1}\dfrac{\tilde\delta}{\xi-1+\delta\left(1-1/h\right)}.$
Let us assume that (\ref{cnec1}) holds. Then, in view of 
$\frac{\partial\texttt I^c_k}{\partial k^2}=2\delta_1k^2+\delta_2,$
it easily follows that $\texttt I^c_k$ reaches its minimum at 
$
\bar k^2_{min}=-\frac{\delta_2}{2\delta_1}$
and is given by 
\begin{equation}\label{Icmin}
(\texttt I^c_k)_{min}=-\frac{\delta^2_2}{4\delta_1}+\delta_3=-\frac{\left(\tilde\delta-\mu_1 d I_c\right)^2}{4\delta_1}+\texttt I_2.
\end{equation}
Simple calculation shows that $(\texttt I^c_k)_{min}$ is negative if and only if
\begin{equation}
\mu^2_1 I_c^2 d^2-2I_c\mu_1 \tilde\delta d + \tilde\delta^2- 4\delta_1(\mu_1+\mu_2)(\xi+\delta/h)>0\end{equation}
i.e. if and only if 
\begin{equation}\label{sufftur}
d>\frac{\tilde\delta+2\sqrt{\delta_1 (\mu_1+\mu_2)(\xi+\delta/h)}}{\mu_1I_c}:=d_c.\end{equation}
Then, in view of (\ref{cnec1}), the condition (\ref{sufftur}) is sufficient for the occurrence of Turing instability. Collecting the obtained results, we proved that: \emph{when $R_0>1$ and $d>d_c$, then $E_c$, stable in the absence of self diffusion and chemotaxis, becomes unstable when spatial variation is allowed. In this case, Turing patterns emerge.}

These analytical results are in agreement with those ones found by the numerical simulations. The instability threshold $d_c$ is strictly dependent on the model parameters. In particular, as $\delta$ tends to zero, one can recover the instability threshold in the absence of the logistic growth of $T$ cells (that is the case analyzed in \cite{Stancevic}). This is given by
\begin{equation}\label{dcm}
d_M=\lim_{\delta\to 0}d_c=\frac{\eta\left\{(\mu_1+\mu_2)d_1+\mu_2 d_1+\mu_3 d_3+\xi(d_2+d_3)+2\sqrt{\xi\delta_1(\mu_1+\mu_2)}\right\}}{\mu_1(\xi-1)}.\end{equation}
Numerical simulations show that, as the parameters vary in their biological range, $d_c$ is a decreasing function of $\delta$ and $h$ (see Fig. 5). 

\begin{figure}[htbp]
    \centering
    \includegraphics[width=0.4\linewidth]{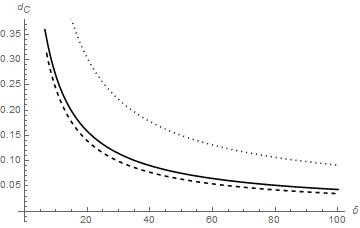}
\caption{Asymptotic behaviour of $d_c$ with respect to $\delta$ for different values of $k$. All the parameters are set as in Sect. 2, while $k=1000$ (dashed line) , $500$ (continuous line). Dotted line represents the AIDS case where $k=180$ and $s=5$.}
\end{figure}

Adopting a logistic growth model for $CD4^+T$ cells allows for a more timely observation of hotspots in infections. This means that the logistic model, which accounts for limiting factors that eventually slow down growth as the population reaches its carrying capacity, provides a more accurate representation of $T$ cell dynamics in the context of an immune response.

\section{Weakly nonlinear analysis}

In this section, we conduct a weakly nonlinear analysis to explore solution behavior beyond the Turing instability threshold. Unlike linear analysis, which cannot predict pattern amplitudes or shapes accurately once instability occurs, weakly nonlinear analysis incorporates crucial nonlinear effects. This approach involves three main steps: 1) expanding perturbations using a small parameter to measure distance from the bifurcation point; 2) deriving expanded equations that include nonlinear terms up to a specified order, thereby approximating the dynamics of perturbation amplitudes; and 3) examining solution behavior, identifying stationary solutions of these equations, and analyzing their stability properties. This analysis is essential for understanding how patterns stabilize and for predicting the characteristics of stable patterns that emerge.\\
\noindent
Denoting by $U=u_1-1; V=u_2-1; W=u_3-1$ the perturbation fields to the coexistence equilibrium $E_c=(1,1,1)$, the nonlinear system governing the evolution of $\mathbf U=(U,V,W)^T$ is
\begin{equation}\label{nonline}
\frac{\partial\mathbf U}{\partial t}=\mathcal L(d)\mathbf U+\mathcal N_1(\mathbf U)-d I_c\left(\begin{array}{c}
                                                                                         \nabla U\cdot\nabla V+U\Delta V\\
                                                                                         0\\
                                                                                         0
                                                                                        \end{array}
\right),\end{equation}
with $\mathcal L(d)=\mathcal L_c+\mathcal D_c\Delta$ is the linear operator, $\mathcal N_1(\mathbf U)$ is the nonlinear operator given by
$$
\mathcal L(d)=\left(\begin{array}{ccc}
                     a_{11}+d_1\Delta&-dI_c\Delta&a_{13}\\
                     a_{21}&a_{22}+d_2\Delta&a_{23}\\
                     0&a_{32}&a_{33}+d_3\Delta
                    \end{array}\right),\quad \mathcal N_1(\mathbf U)=\left(\begin{array}{c}                                                    -\delta/h U^2-\eta I_c UW\\
                                                                 \mu_1 UW\\
                                                0                                                                \end{array}\right).
$$

Let $\varepsilon^2=(d-d_c)/d_c$ be a small parameter representing the dimensionless distance from the threshold. Introducing a new scale time coordinate $T=\varepsilon^2 t$, in such a way that $\partial_t\to \partial t+\varepsilon^2 \partial_T$, let us expand $d$ and $\mathbf U$ such that
$$
 d=d_c+\varepsilon^2 d^{(2)}+o(\varepsilon^4),\quad
 \mathbf U=\varepsilon\mathbf u_1+\varepsilon^2 \mathbf u_2+\varepsilon^3\mathbf u_3+o(\varepsilon^4),
$$
with $\mathbf u_i=(u_i,v_i,w_i)\, (i=1,2,3)$.
The operator $\mathcal L(d)$ can be expanded as
$$
\mathcal L(d)=\mathcal L_{dc}+\varepsilon^2\left(\begin{array}{ccc}
                              0&-d^{(2)}I_c\Delta &0\\
                              0&0&0\\
                              0&0&0
                                                 \end{array}
\right)+o(\varepsilon^4),
$$
with $\mathcal L_{dc}=
\left(\begin{array}{ccc}
                     a_{11}+d_1\Delta&-d_cI_c\Delta&a_{13}\\
                     a_{21}&a_{22}+d_2\Delta&a_{23}\\
                     0&a_{32}&a_{33}+d_3\Delta
                    \end{array}\right)
$. 
Substituting the previous expansion into (\ref{nonline}), and collecting the coefficients of the same order in $\varepsilon$, one gets:
\begin{itemize}
 \item at $o(\varepsilon)$
\begin{equation}\label{ordine1}
\mathcal L_{dc}\mathbf u_1=\mathbf 0;\end{equation}
\item at $o(\varepsilon^2)$
\begin{equation}\label{ordine2}
\mathcal L_{dc}\mathbf u_2=d_cI_c\left(\begin{array}{c}
                  \nabla u_1\cdot\nabla v_1+u_1\Delta v_1\\
                  0\\
                  0
                                                 \end{array}\right)+\left(\begin{array}{c}
               \delta u^2_1/h+\eta I_cu_1w_1\\
               -\mu_1u_1w_1\\
               0
                                \end{array}\right);
\end{equation}
\item at $o(\varepsilon^3)$
\begin{equation}\label{ordine3}\begin{array}{ll}
\mathcal L_{dc}\mathbf u_3&=\frac{\partial\mathbf u_1}{\partial T}
+d^{(2)}\left(\begin{array}{c}I_c\Delta v_1\\
               0\\
               0\end{array}\right)

+\left(\begin{array}{c}
                    2\delta u_1u_2/h+\eta I_cu_2w_1+\eta I_cu_1w_2\\
                    -\mu_1u_2w_1-\mu_1u_1w_2\\
                    0
                    \end{array}\right)\\
                    &+ d_cI_c\left(\begin{array}{c}
                     \nabla u_2\cdot\nabla v_1+\nabla u_1\cdot\nabla v_2+u_2\Delta v_1+u_1\Delta v_2\\
                     0\\
                     0
                    \end{array}\right).
                    \end{array}\end{equation}
 \end{itemize}
Let us solve (\ref{ordine1}). In view of the boundary conditions, one can look for solutions of type
\begin{equation}
\mathbf u_1= A(T) \bm{ \rho} \cos (k_c x),
\end{equation}
being $ \bm{\rho}=(\rho_u, \rho_v, \rho_w)^T$, $A(T)$ is the amplitude of pattern. Solving (\ref{ordine1}), the vector $\bm{\rho}$ can be normalized as
\begin{equation}\label{rho}
\bm{\rho}=\left(\begin{array}{c}
\frac{-a_{23}a_{32}+(d_2k^2_c-a_{22})(d_3k^2_c-a_{33})}{a_{21}a_{32}}\\
(d_3k^2_c-a_{33})/a_{32}\\
1
                   \end{array}\right).\end{equation}
 Substituting in (\ref{ordine2}), one has 
 \begin{equation}\label{ordine2*}
 \begin{array}{ll}
 \mathcal L_{dc}\left(\begin{array}{c}
                   u_2\\
                   v_2\\
                   w_2
                  \end{array}\right)&
                  =d_cI_ck^2_cA^2\left(\begin{array}{c}
                  \rho_u\rho_v\sin^2(k_cx)-\cos^2(k_cx)\\
                  0\\
                  0
                                                 \end{array}\right)
                                                 \\&+
                                                 \left(\begin{array}{c}
               \delta \rho^2_u/h+\eta I_c\rho_u\rho_w\\
               -\mu_1\rho_u\rho_w\\
               0
                                \end{array}\right)
                          A^2\cos^2(k_cx):=\left(\begin{array}{c}
                                                  F_u\\
                                                  F_v\\
                                                  F_w
                                                 \end{array}
\right)\end{array} \end{equation}
Denoting by $\mathbf \Psi=(\Psi_1,\Psi_2,\Psi_3)^T$,  the eigenvector of the null eigenvalue of the adjoint operator of $\mathcal L(d)$, given by 
\begin{equation}\label{psi}
\mathbf \Psi=\left(\begin{array}{c}
                    a_{21}/(d_1k^2_c-a_{11})\\
                    1\\
                    \left[ a_{13}a_{21}+a_{23}(d_1k^2_c-a_{11})\right]/\left[(d_1k^2_c-a_{11})(d_3k^2_c-a_{33})\right]
                   \end{array}
\right)\end{equation}
the Fredholm solvability condition  $<\mathbf F, \mathbf \Psi>=0$ with $<\cdot,\cdot>$ the scalar product in $L^2(0,2\pi/k_c)$, $\mathbf F=(F_u,F_v,F_w)^T$, is satisfied.
Then, in view of the linearity of the problem, the solution of (\ref{ordine2*}) can be written as
\begin{equation}\label{solordine2}
\mathbf u_2=A^2(\mathbf u_{20}+\mathbf u_{22}\cos (2k_c x)),\end{equation}
with $\mathbf u_{20}=(\mu_u,\mu_v,\mu_w), \mathbf u_{22}=(\theta_u,\theta_v,\theta_w)$ solutions of 
\begin{equation}
\begin{cases}
 (a_{11}\!-\!d_1k^2_c)(\mu_u\!+\!\theta_u)\!+\!I_cd_ck^2_c(\mu_v\!+\!\theta_v)\!+\!a_{13}(\mu_w\!+\!\theta_w)\!=\!-d_cI_ck^2_c\rho_u\rho_v\!+\!(\delta/h) \rho^2_u\!+\!\eta I_c\rho_u\rho_w,\\
 (a_{11}-d_1k^2_c)(\mu_u-\theta_u)+I_cd_ck^2_c(\mu_v-\theta_v)+a_{13}(\mu_w-\theta_w)=d_cI_ck^2_c\rho_u\rho_v,\\
 a_{21}(\mu_u+\theta_u)+(a_{22}-d_2k^2_c)(\mu_v+\theta_v)+a_{23}(\mu_w+\theta_w)=-\mu_1\rho_u\rho_w,\\
 a_{21}(\mu_u-\theta_u)+(a_{22}-d_2k^2_c)(\mu_v-\theta_v)+a_{23}(\mu_w-\theta_w)=0,\\
 a_{32}(\mu_v+\theta_v)+(a_{33}-d_3k^2_c)(\mu_w+\theta_w)=0,\\
 a_{32}(\mu_v-\theta_v)+(a_{33}-d_3k^2_c)(\mu_w-\theta_w)=0.
\end{cases}
\end{equation}
Substituting into (\ref{ordine3}), one has that
\begin{equation}\label{ordine3*}
\mathcal L_{dc}\mathbf u_3=\left(\frac{dA}{dT}\bm{\rho}+A\mathbf G^{(1)}+A^3\mathbf G^{(2)}\right)\cos(k_cx)+A^3\mathbf G^{(*)}\cos(3k_cx),\end{equation}
                  with 
   $$
   \mathbf G^{(1)}=-d^{(2)}k^2_cI_c\left(\begin{array}{c}
                                          \rho_v\\
                                          0\\
                                          0
                                         \end{array}
\right);
   $$
   $$
   \begin{array}{ll}
   \mathbf G^{(2)}&=d_cI_ck^2_c\left(\begin{array}{c}
                                \rho_v\theta_u+\rho_u\theta_v-\rho_v\mu_u-\theta_u\rho_v/2+\rho_u\theta_v\\
                                0\\
                                0
                               \end{array}
\right)\\
&+
\left(\begin{array}{c}
       2\delta \rho_u\mu_u/h+\delta\rho_u\theta_u/h+\eta I_c\rho_w \mu_u+\eta I_c\rho_w\theta_u/2+\eta I_c\rho_u\rho_w+\eta I_c\rho_u\theta_w/2\\
       -\mu_1\rho_w\mu_u-\mu_1\rho_w\theta_u/2-\mu_1\rho_u\mu_w-\mu_1\rho_u\theta_w/2\\
       0
      \end{array}
\right)
\end{array}
   $$
    $$
    \mathbf G^{(*)}=-d_cI_ck^2_c\left(
    \begin{array}{c}
     \rho_v\theta_u+\rho_u\theta_v+\theta_u\rho_v/2+\rho_u\theta_v\\
     0\\
     0
    \end{array}
    \right)+\left(
    \begin{array}{c}
     \delta \rho_u\theta_u/h+\eta I_c\rho_w\theta_u/2+\eta I_c\rho_u\theta_w/2\\
     -\mu_1(\rho_w\theta_u+\rho_u\theta_w)/2\\
     0
    \end{array}
    \right).
    $$
Then, the Fredholm solvability condition for equation (\ref{ordine3*}) leads to the Stuart-Landau equation for the amplitude $A(T)$:
   \begin{equation}\label{SL}
   \frac{dA}{dT}=\sigma A-LA^3,\end{equation}
   where
   \begin{equation}\label{sigmaL}
   \sigma=-\frac{(\mathbf G^{(1)},\mathbf \Psi)}{(\bm{\rho},\mathbf \Psi)},\quad L=\frac{(\mathbf G^{(2)},\mathbf \Psi)}{(\bm{\rho},\mathbf \Psi)}.\end{equation}
Straightforward calculation shows that, at the onset of instability (i.e. $d>d_c$), $\sigma>0$. The Landau constant $L$ can be positive or negative. 
\begin{itemize}
\item[1)]\emph{The supercritical case}.
When $L>0$ there is a supercritical bifurcation. The Stuart-Landau equation (\ref{SL}) admits as stable equilibrium $A_\infty=\sqrt{\sigma/L}$ and the solution reverts, as time tends to infinity, to
\begin{equation}\label{supbif}
\mathbf U=\varepsilon\bm{\rho} \sqrt{\sigma/L} \cos(k_cx)+\varepsilon^2 \sigma/L (\mathbf u_{20}+\mathbf u_{22}\cos(2k_cx))+o(\varepsilon^3).\end{equation}
\item[2)]\emph{The subcritical case}.
When $L<0$ there is a subcritical bifurcation. The weakly nonlinear analysis has to be extended up to the fifth order. Introducing the multiple time scales $T$ and $T_1$ as follows
$
t=\frac{T}{\varepsilon^2}+\frac{T_1}{\varepsilon^4}+\cdots,$
and expanding $d$ and $\mathbf U$ up to the fifth order in $\epsilon$, one has the same equations up to the third order given by (\ref{ordine1})-(\ref{ordine3}). Then, at the order three, one recovers (\ref{SL}) with $A=A(T,T_1)$ at the place of $A=A(T)$ and the time derivative is a partial derivative with respect to $T$. If the solvability condition $<\mathbf G,\mathbf\Psi>=0$ is satisfied, then the solution is $\mathbf u_3=(u_3,v_3,w_3)^T$ given by 
\begin{equation}\label{sol3}
\mathbf u_3=(A\mathbf u_{31}+A^3 \mathbf u_{32})\cos (k_cx)+A^3\mathbf u_{33}\cos (3k_cx),\end{equation}
where $\mathbf u_{3i}, (i=1,2,3)$ are solutions of
$$
\begin{array}{l}
(\mathcal L_{c}-k^2_cD_c)\mathbf u_{31}=\sigma\bm{\rho}+\mathbf G^{(1)},\\
(\mathcal L_{c}-k^2_cD_c)\mathbf u_{32}=-L\bm{\rho}+\mathbf G^{(2)},\\
(\mathcal L_{c}-9k^2_cD_c)\mathbf u_{33}=\mathbf G^{(*)}
\end{array}
$$
At $o(\varepsilon^4)$, one has that the solvability condition is verified and the solution can be written as
$$
\mathbf u_4=A^2\mathbf u_{40}+A^4\mathbf u_{41}+(A^2\mathbf u_{42}+A^4\mathbf u_{43})\cos (2k_cx)+A^4 \mathbf u_{44}\cos (4k_cx),
$$
where $\mathbf u_{4i}, (i=0,\cdots, 4)$ are obtained by solving five linear systems (their expressions are quite intricate, so we will not include them here).
\\
\noindent
Finally, on balancing the coefficients in $\varepsilon^5$, the Fredholm solvability condition leads to the Stuart-Landau equation
\begin{equation}\label{SL2}
\frac{\partial A}{\partial T_1}=\tilde\sigma A- \tilde L A^3+\tilde Q A^5,\end{equation}
where $\tilde \sigma, \tilde L, \tilde Q$ depend on the system parameters. Adding (\ref{SL}) and (\ref{SL2}), one obtains the quintic Stuart-Landau equation
\begin{equation}\label{SL3}
\frac{\partial A}{\partial T}=\bar\sigma A-\bar L A^3+\bar Q A^5,\end{equation}
being $\bar \sigma=\sigma+\varepsilon^2\tilde\sigma$. When $\bar\sigma>0,\bar L<0, \bar Q<0$, there exists a real stable equilibrium $A_{\infty}=\sqrt{\frac{\bar L-\sqrt{\bar L^2-4\bar\sigma \bar Q}}{2\bar Q}}$ representing the asymptotic value of the amplitude. 
\end{itemize}

\section{Conclusion}

This paper introduces a reaction-diffusion mathematical model for the early stages of HIV infection. Experimental evidence (referenced in \cite{Miller2003}, \cite{Lin2006}) supports the inclusion of random diffusion and chemotaxis in the model to capture biological dynamics. A novel addition is the incorporation of logistic growth for CD4+ T cells.
The model identifies two stable equilibria: the disease-free equilibrium (DFE), where all CD4+ T cells are healthy, and the endemic equilibrium (EE), where all components are present. The DFE always exists, while the EE emerges when the DFE becomes unstable, contingent on a sufficient carrying capacity of T cells.
Preliminary numerical analysis indicates that exceeding a specific threshold in the dimensionless chemotactic term leads to spatial pattern formation. The study establishes necessary and sufficient conditions for the linear stability of both equilibria, under homogeneous and heterogeneous scenarios. These conditions are crucial for understanding long-term disease dynamics.
Furthermore, the study determines closed-form conditions for Turing instability, indicating diffusion-driven instability occurs at a lower threshold compared to models without logistic growth of T cells. This suggests the logistic growth model facilitates earlier detection of infection hot spots.
Finally, a weakly nonlinear analysis is conducted to delve into pattern formation and determine their amplitudes, providing deeper insights into the model's predictions.

\bibliographystyle{unsrt}
\bibliography{References}

\end{document}